\documentclass{aa}

\usepackage{soul}
\usepackage{graphicx, textcomp}
\usepackage{multirow}
\usepackage{natbib}
\usepackage[switch]{lineno}
\usepackage[T1]{fontenc}
\usepackage{lmodern}
\usepackage{placeins}
\usepackage{caption}

\usepackage{txfonts}

 \usepackage{hyperref}
 \hypersetup{
      colorlinks   = true,
      citecolor    = blue
 }

\begin{document}

   \title{The twin-jet system in the FRII radio galaxy 3C 452}
   \subtitle{A sub-parsec scale VLBI study}

\author{E. Madika\inst{1,2}
\and B. Boccardi\inst{1}
\and L. Ricci\inst{1,3}
\and P. Grandi\inst{4}
\and E. Torresi\inst{4}
\and G. Giovannini\inst{5,6}
\and M. Kadler\inst{3}
\and J. A. Zensus\inst{1}
}

\institute{
Max-Planck-Institut f\"{u}r Radioastronomie, Auf dem H\"{u}gel 69, D-53121 Bonn, Germany 
\and Leiden Observatory, Leiden University, PO Box 9513, 2300 RA Leiden, The Netherlands
\and Julius-Maximilians-Universit{\"a}t W{\"u}rzburg, Fakult{\"a}t für Physik und Astronomie, Institut für Theoretische Physik und Astrophysik, Lehrstuhl für Astronomie, Emil-Fischer-Str. 31, D-97074 Würzburg, Germany
\and INAF – Osservatorio di Astrofisica e Scienza dello Spazio di Bologna, Via Gobetti 101, 40129 Bologna, Italy
\and INAF–Istituto di Radioastronomia, Bologna, Via Gobetti 101, 40129 Bologna, Italy
\and Dipartimento di Fisica e Astronomia, Università degli Studi di Bologna, Via Gobetti 93/2, 40129 Bologna, Italy
}

   \date{Received 20 October 2025 / Accepted 25 February 2026}

  \abstract{}{We present a comprehensive multifrequency VLBI analysis of the FRII, high-excitation radio galaxy 3C 452, aiming to resolve and analyze for the first time its twin-jet structure on sub-parsec scales.}{Our data set comprises high-sensitivity array (HSA) observations at 4.9, 8.4, 15.4, 23.6, and 43.2 GHz. 
Through fitting methods performed in both the visibility and the image plane, we were able to trace the jet expansion from scales of a few thousand to nearly $10^5$ Schwarzschild radii ($R_{\rm S}$), on both the approaching and receding jets. Additionally, we derived the core brightness temperatures and Doppler factors to constrain the jet's orientation and intrinsic speed.}
{Our study provides the first detailed description of the twin-jet system in 3C 452 on VLBI scales, confirming it as a rare FRII source with jets detected down to millimeter wavelengths. We resolved both jet and counter-jet down to scales of a few thousand $R_\mathrm{S}$, revealing a symmetric, parabolically expanding structure with power-law indices $k\approx 0.66$ (jet) and $k\approx 0.47$ (counter-jet). Jet-to-counter-jet intensity ratios remain nearly constant on larger scales but rise sharply within the inner 1–1.5 mas, as revealed by the higher-frequency data, consistent with increasing jet speed. The brightness temperature analysis yields low Doppler factors ($\delta \sim 0.03$–$0.83$), indicative of Doppler de-boosting due to the large viewing ($\theta \approx 70^\circ$) and/or a magnetically dominated jet base.  A spectral index analysis reveals a strongly inverted core spectrum ($\alpha > 2$) with additional absorption at the highest frequencies, followed by a sharp steepening ($\alpha\sim-2.5$) to optically thin values in the innermost jet. Finally, a comparison between broad- and narrow-line high-excitation radio galaxies shows that jets in narrow-line sources, such as 3C 452 and Cygnus A, complete collimation at $\lesssim 10^{5} R_{\mathrm{S}}$, whereas broad-line sources exhibit shape transitions at $10^{6}$–$10^{7} R_{\mathrm{S}}$, suggesting that orientation plays an important role in the observed collimation scales.}
  {}

\titlerunning{The twin jet system in the FRII radio galaxy 3C 452}
\authorrunning{E. Madika et al.}

   \maketitle

\section{Introduction}
Extragalactic jets are collimated outflows of relativistic plasma powered by the accretion of material onto a supermassive black hole at the center of active galactic nuclei (AGN) \citep[see, e.g.,][]{Blandford2019}. High-frequency very long-baseline interferometry (VLBI) observations are a unique tool to unveil the fundamental processes that drive their formation \citep[see][and references therein]{Boccardi2017}. Most VLBI studies of AGN jets have focused on blazars, while radio galaxies offer a unique advantage because projection effects and Doppler boosting play a minor role compared to blazars. In particular, studies of nearby radio galaxies with jets oriented at a large viewing angle can be highly informative because they allow us to probe the intrinsic jet properties in the immediate vicinity of the black hole.

Radio galaxies known to be bright enough for probing the jet sub-parsec scales down to millimeter (mm) wavelengths are rare, especially among the class of powerful high-excitation galaxies (HEGs) \citep{2025A&A...695A.118B}. These sources are thought to be powered by radiatively efficient, cold accretion disks fed by cold gas \citep{HB2014}, and typically produce powerful jets developing Fanaroff-Riley II (FRII) morphologies \citep{fanaroff} on the kiloparsec (kpc) scales. Even rarer are HEGs showing sufficiently bright two-sided jets on such small scales, a property that enables important tests of the jet symmetry to be performed. The two-sided FRII jet in Cygnus A is currently the only case in which such mm-VLBI studies could be carried out \citep{boccardi15}. With the aim of identifying new targets for jet formation studies, including HEGs, \cite{2025A&A...695A.118B} have carried out a VLBI experiment using the High Sensitivity Array (HSA) at 22 GHz (1 cm) and 43 GHz (7 mm), considering 16 radio galaxies poorly explored on VLBI-scales. These were selected among those characterized by high spatial resolution in units of Schwarzschild radii, which resulted from their proximity ($z<0.1$) and large black hole mass ($\log {M_{\rm BH}}>8.5$). 

Among the HEGs in this sample, the radio galaxy 3C\,452 emerged as the most interesting target.  
This source is located at redshift $z=0.081$ and hosts a black hole with mass $M_\mathrm{BH} = {\sim}8 \times 10^8 M_\odot$ \citep{Koss2022}. 
The existence of a radiatively efficient accretion disk at the center of this object is indicated by optical spectroscopic studies \citep[e.g.][]{Buttiglione2010}, which also classify 3C\,452 as a high-excitation galaxy (HEG) based on the Excitation Index (EI) parameter.\footnote{The Excitation Index is defined as 
$\mathrm{EI} = \log [\mathrm{O\,III}]/\mathrm{H}\beta 
- 1/3\left( \log [\mathrm{N\,II}]/\mathrm{H}\alpha 
+ \log [\mathrm{S\,II}]/\mathrm{H}\alpha 
+ \log [\mathrm{O\,I}]/\mathrm{H}\alpha \right)$ \citep{Buttiglione2010}.} 
 The lack of broad emission lines, that is, the source classification as a narrow-line radio galaxy \citep[e.g.,][]{veron}, implies obscuration of the nuclear regions due to presence of a compact absorber, as confirmed by X-ray observations \citep{fioretti}.
In large-scale radio observations performed with the Very Large Array (VLA), 3C\,452 exhibits a symmetric double morphology  \citep[e.g.,][]{black}. Its largest angular size of $\sim 5$ arcmin in 1.4 GHz maps corresponds to a projected linear size of $\sim 450$ kpc at the source redshift of $z=0.081$ \citep[e.g.,][]{Shelton}. In addition, low-frequency observations have revealed the presence of Mpc-scale relic radio lobes associated with this source \citep{Sirothia}. The source was previously almost unexplored on VLBI scales. A single image at 5 GHz was presented by \cite{giovannini}. This revealed a highly symmetric two-sided structure with a total flux density of ${\sim}130$\,$\rm mJy$. Taking into account the core dominance and jet symmetry, the authors constrained the jet to be oriented at an angle $\theta \geq 60^{\circ}$ with respect to the line of sight. The 22 GHz and 43 GHz observations by \cite{2025A&A...695A.118B} detected a twin-jet structure down to the sub-parsec scales. At these high frequencies, 3C\,452 was shown to be still relatively bright, with a total flux density of ${\sim}68$\,$\rm mJy$ at 22 GHz and ${\sim}53$\,$\rm mJy$ at 43 GHz. Therefore, follow-up VLBI observations were requested with the aim of addressing several open questions concerning the jet launching regions in powerful sources.

In this article, we present VLBI observations of 3C\,452 at frequencies between 5 GHz and 43 GHz. The aim of this work is to obtain a first description of the fundamental parameters of this jet on sub-parsec scales (i.e., orientation and Doppler factor) and to investigate whether the physical processes of jet acceleration and collimation are still taking place on the examined scales.
At the source redshift ($z = 0.0811$), and assuming a $\Lambda$CDM cosmology with $H_0 = 71\,\mathrm{km\,s^{-1}\,Mpc^{-1}}$, $\Omega_{\mathrm{M}} = 0.27$, and $\Omega_{\Lambda} = 0.73$, an angular scale of $1\,\mathrm{mas}$ corresponds to $\approx 1.508\,\mathrm{pc}$. Given a black hole mass of $8 \times 10^8\,M_{\odot}$, this translates to $\approx 19{,}738$ Schwarzschild radii ($R_{\mathrm{S}}$) per mas. Thus, these observations provide an angular resolution down to a few thousand $R_{\rm S}$, sufficient to resolve the jet base at a close distance from the central engine.

\begin{table*}[!ht]
\centering
\footnotesize
\caption{Log of observations and characteristics of the clean maps forming the multi-frequency data-set. }
\label{tab:table 1}
\begin{tabular}{cccccccccc}
\hline
\hline
Date & P.C. & Array & Freq. & Beam & Eq. Beam & S$_{\mathrm{peak}}$ & S$_{\mathrm{total}}$ & rms \\
 & & & $\mathrm{[GHz]}$ & $\mathrm{[mas\times mas, deg]}$ & $\mathrm{[mas]}$ & $\mathrm{[mJy/beam]}$ & $\mathrm{[mJy]}$ & $\mathrm{[mJy/beam]}$ \\
\hline
\multicolumn{9}{c}{Uniform weighting} \\
\hline
2022-01-02 & BM516A & VLBA+EB & 4.9 & $2.06\times 1.29, -5^{\circ}$ & $1.63$ & $21$ & $97$ & $0.03$ \\
 & & & 8.4 & $1.17\times 0.61, -13^{\circ}$ & $0.84$ & $19$ & $89$ & $0.05$ \\
 & & & 15.4 & $0.52\times 0.33, -15^{\circ}$ & $0.41$ & $22$ & $69$ & $0.05$ \\
 & & & 23.6 & $0.45\times 0.29, -16^{\circ}$ & $0.36$ & $26$ & $61$ & $0.09$ \\
\hline
2022-08-06 & BM516B & VLBA+EB & 4.9 & $1.78\times 1.00, -17^{\circ}$ & 1.33 & 19 & 100 & 0.02 \\
 & & & 8.4 & $1.09\times 0.58, -16^{\circ}$ & 0.79 & 19 & 89 & 0.03 \\
 & & & 15.4 & $0.68\times 0.35, -10^{\circ}$ & 0.49 & 25 & 72 & 0.05 \\
 & & & 23.6 & $0.46\times 0.21, -13^{\circ}$ & 0.31 & 22 & 63 & 0.06 \\
 & & & 43.2 & $0.22\times 0.10, -18^{\circ}$ & 0.15 & 28 & 56 & 0.06 \\
\hline
\multicolumn{9}{c}{Natural weighting} \\
\hline
2022-01-02 & BM516A & VLBA+EB & 4.9 & $2.57\times 1.76, -0.4^{\circ}$ & $2.13$ & $25$ & $97$ & $0.02$ \\
 & & & 8.4 & $1.58\times 0.88, -10^{\circ}$ & $1.18$ & $23$ & $89$ & $0.04$ \\
 & & & 15.4 & $0.71\times 0.47, -9^{\circ}$ & $0.58$ & $26$ & $70$ & $0.04$ \\
 & & & 23.6 & $0.66\times 0.56, -17^{\circ}$ & $0.61$ & $29$ & $61$ & $0.04$ \\
\hline
2022-08-06 & BM516B & VLBA+EB & 4.9 & $2.13\times 1.23, -16^{\circ}$ & 1.62 & 22 & 100 & 0.02 \\
 & & & 8.4 & $1.37\times 0.76, -15^{\circ}$ & 1.02 & 23 & 89 & 0.02 \\
 & & & 15.4 & $0.87\times 0.50, -8^{\circ}$ & 0.66 & 28 & 72 & 0.03 \\
 & & & 23.6 & $0.57\times 0.31, -14^{\circ}$ & 0.42 & 26 & 63 & 0.04 \\
 & & & 43.2 & $0.28\times 0.13, -17^{\circ}$ & 0.19 & 31 & 56 & 0.04 \\
\hline
\hline
\end{tabular}

\tablefoot{
Col. 1: Date of observation. Col. 2: Project code. Col. 3: Array. Col. 4: Frequency. 
Col. 5: Beam FWHM and position angle. Col. 6: Equivalent circular beam. Col. 7: Peak intensity. 
Col. 8: Total flux density.
Col. 9: Image noise. Values are given for untapered data with both uniform and natural weighting.}

\end{table*}

The paper is structured as follows. In Sect.\,\ref{sec:data set}, we present the multi-frequency VLBI observations, describe the data reduction process, and outline the analysis methods, including model fitting, image alignment across frequencies, and the construction of stacked maps. The results and their interpretation are presented in Sect.\,\ref{sec:results}, covering the flux density and jet-to-counter-jet profiles, the core brightness temperatures, the jet collimation properties and the spectral index distribution along the ridge line. A comparison with Cygnus A and other HEGs is presented in Sect.\,\ref{sec:comparison}. We summarize our conclusions in Sect.\,\ref{sec:conclusions}.

\begin{table*}[!ht]
    \centering
    \caption{Results of the 2D cross-correlation for core shift determination. }
    \small
    \label{tab:table 2}
    \begin{tabular}{cccccccc}
        \hline
        Date & P.C. & Freq. pair & Common Beam & $\Delta x$, $\Delta y$ & $\Delta z=\sqrt{{\Delta x}^2+{\Delta y}^2}$ & Uncertainty & Correlation \\
        & & [GHz] & [mas $\times$ mas] & [mas, mas] & [mas] & [mas] & Coefficient \\
        \hline
        2022-01-02 & BM516A & 4.9/8.4 & $1.66\times1.66$ & $0.68, 0.00$ & $0.68$ & $0.23$ & 0.999\\
        & & 8.4/15.4 & $0.88\times0.88$ & $0.45, 0.00$ & $0.45$ & $0.12$ & 0.979\\
        & & 15.4/23.6 & $0.60\times0.60$ & $0.24, 0.00$ & $0.24$ & $0.08$ & 0.938\\
        \hline
        2022-08-06 & BM516B & 4.9/8.4 & $1.32\times1.32$ & $0.65, 0.00$ & $0.65$ & $0.19$ & 0.989\\
        & & 8.4/15.4 & $0.84\times0.84$ & $0.32, 0.00$ & $0.32$ & $0.12$ & 0.979\\
        & & 15.4/23.6 & $0.54\times0.54$ & $0.20, 0.05$ & $0.21$ & $0.08$ & 0.961\\
        & & 23.6/43.2 & $0.30\times0.30$ & $0.39, 0.00$ & $0.39$ & $0.04$ & 0.961\\
        \hline
    \end{tabular}
    \tablefoot{Col. 1: Observation date. Col. 2: Project code. Col. 3: Frequency pair. Col. 4: Common circular restoring beam. Col. 5: Shifts in RA and Dec. Col. 6: Shift magnitude. Col. 7: Positional uncertainty. Col. 8: Correlation coefficient.
}
\end{table*}

\section{Data set and analysis}

\label{sec:data set}
\begin{figure}[!h]
    \centering
        \includegraphics[width=0.5\textwidth]{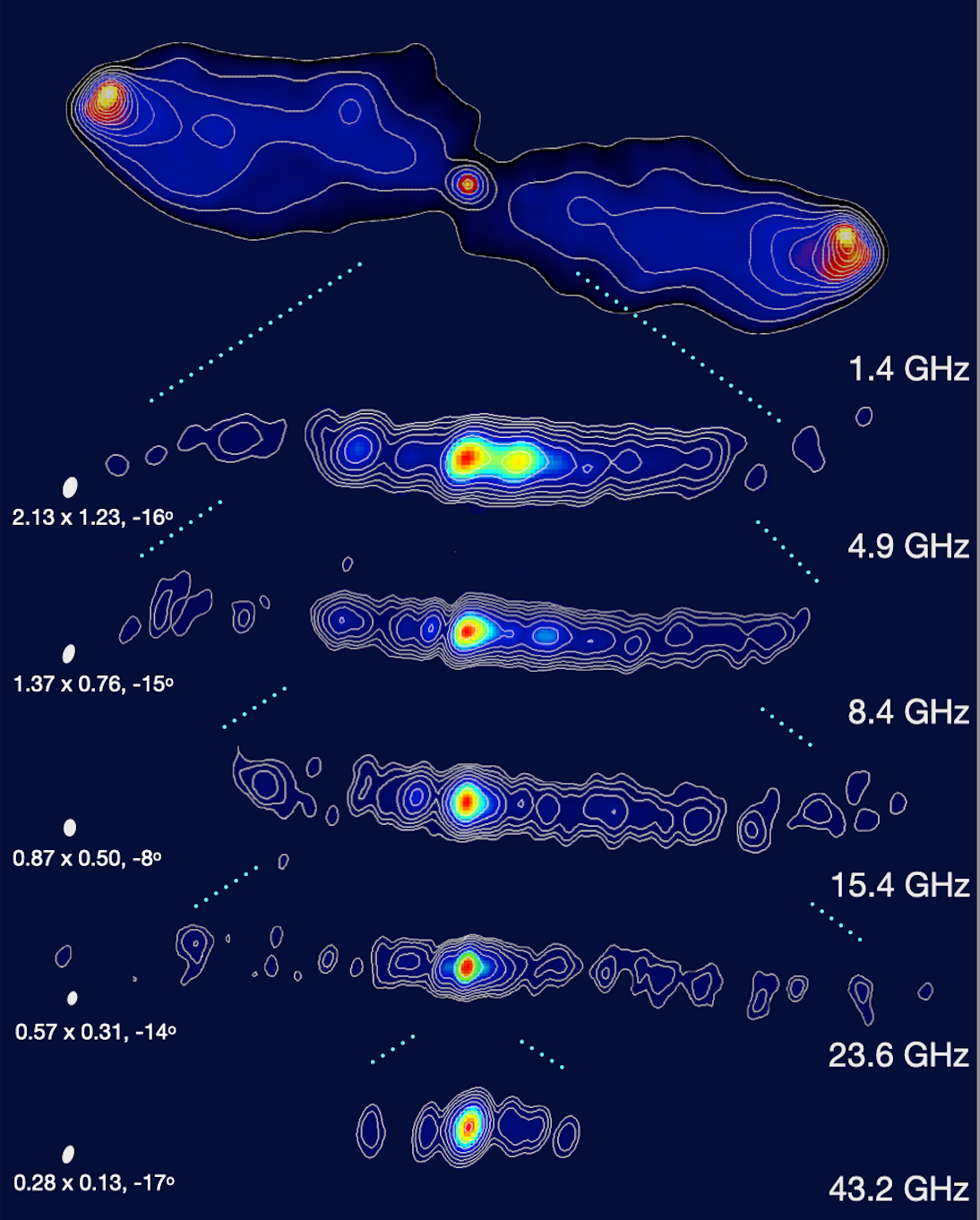}
       \caption{The twin-jet structure of 3C 452 from kiloparsec to sub-parsec scales. The top image was obtained from VLA data at 1.4 GHz \citep{Leahy1996_Atlas}. The following maps show the VLBI structure at different frequencies
(4.9 GHz, 8.4 GHz, 15.4 GHz, 23.6 GHz, 43.2 GHz) based on the analyzed data of our project code BM516B
presented in Sect.~\ref{sec:data set}. Images were produced using natural weighting. The maps have been aligned to the central brightest feature. The complete series of images considered
in the article is presented in the Appendix~\ref{app:appendix}.}
        \label{fig:big-pic}
\end{figure}

On 2 January 2022, we conducted HSA observations using the Very Long Baseline Array (VLBA) and the Effelsberg 100-m radio telescope (Project code: BM516A). The inclusion of Effelsberg improved angular resolution and sensitivity, necessary for a detailed imaging of the sub-parsec-scale structures in our faint target. The observations spanned a 12-hour block for optimal $(u,v)$-coverage. Data were collected at five frequencies: 4.9 GHz, 8.4 GHz, 15.4 GHz, 23.6 GHz, and 43.2 GHz. A 4 Gbit/s recording mode was used in the experiment. The observations employed 10 VLBA antennas, recording in both polarizations with 256 spectral channels per baseband and a total bandwidth of 128 MHz. The data were correlated using the DiFX software correlator in Socorro with an integration time of 2 seconds. The total on-source times for 3C 452 were 0.800 hours at 4.9 GHz, 1.000 hours at 8.3 GHz, 1.600 hours at 15.3 GHz, 2.092 hours at 23.6 GHz, and 3.867 hours at 43.2 GHz.  
However, at 43.2 GHz, the experiment was affected by severe problems. An error in the correction of azimuth and elevation offsets affected the 7\,mm performance of the VLBA antennas, resulting in a $\sim40\%$  sensitivity loss and a change in the flux density scale. Effelsberg also malfunctioned at 7\,mm. This prevented us from imaging the source at the highest frequency. Consequently, we requested and obtained time to repeat the experiment at all five frequencies (Project code: BM516B). In these observations, which were carried out on August 6, 2022, the Kitt Peak (KP) antenna did not participate due to ongoing power issues after a wildfire. Nevertheless, in this epoch we were able to image the source at all bands. 
These data were calibrated using the Astronomical Image Processing System \citep[AIPS,][]{Greisen} following standard procedures, including an opacity correction at frequencies $\geq 15\,\rm GHz$.
The imaging and self-calibration of the amplitude and phase were performed in \texttt{DIFMAP} \citep[Difference Mapping,][]{Shepherd}. 
The log of observations and the characteristics of the clean images for both epochs are presented in Table~\ref{tab:table 1}. The images are presented in Appendix~\ref{app:appendix}. An overview of the structure of the source from the kiloparsec scale to the sub-parsec scale is shown in Fig.~\ref{fig:big-pic}, including the VLA 1.4 GHz from the 3CRR Atlas \citep{Leahy1996_Atlas} and our natural weighted VLBI images from the project code BM516B at 4.9–43.2 GHz. In Fig.~\ref{fig:big-pic}, we note a misalignment between the sub-parsec and kiloparsec jet position angles.

Such misalignments are common in AGN and may arise from a combination of intrinsic and environmental effects. 
Jet re-orientation can occur when the accretion disk is misaligned with the black hole spin axis, inducing relativistic frame-dragging effects such as Lense–Thirring precession and the associated Bardeen–Petterson warp. These processes can lead to episodic changes in the jet launching direction, particularly if the fueling geometry varies with time. In close binary black hole systems, orbital motion and geodetic precession provide an alternative mechanism that can produce Myr-scale changes in jet orientation, comparable to those inferred from pc–kpc misalignments \citep[e.g.][]{Krause2019}. 
Environmental interactions with the ISM or IGM, as well as jet backflow, may also contribute to observed jet morphology, especially in radio galaxies, whereas parsec-scale misalignments have been discussed primarily in the context of quasars and BL Lac objects \citep[e.g.,][]{Kharb2010}.

\subsection{Model fitting }
\label{sec:model fitting}
With the aim of reconstructing the two-sided jet expansion profile and estimate the core brightness temperature, $T_\mathrm{b}$, at all frequencies, we have modeled the data using the \texttt{MODELFIT} subroutine in \texttt{DIFMAP} by fitting circular Gaussian components to the visibilities, which allowed us to determine, for each feature, the integrated flux density ($S$), radial distance ($r$), position angle ($pa$), and size ($d$), corresponding to the full width at half maximum (FWHM) of the Gaussian. 

The derived \texttt{MODELFIT} component parameters are reported for both projects, BM516A and BM516B, in Table~\ref{tab:table A1} and Table~\ref{tab:table A2} in Appendix~\ref{app:appendix}. 
The radial distance of each feature, reported in column 3, is relative to the position of the most compact component at the jet base, which was set as the origin in each map. The uncertainties for these parameters were estimated following the approach of \cite{2008AJ....136..159L}, based on the signal-to-noise ratio (S/N) of each emission feature. The S/N also defines the resolution limit of each component, which we consider when estimating the brightness temperature of the VLBI cores (Table~\ref{tab:table A3}).

\subsection{Alignment of maps at different frequencies}
\label{sec:alignment}
To meaningfully combine the data across all frequencies, it is essential to refer the measured radial distances of all jet components to a common origin, ideally corresponding to the central supermassive black hole. In each individual map, the origin coincides with the position of the peak emission, which is frequency dependent due to synchrotron opacity effects at the jet base \citep{Blandford79,Lobanov98}.
To correct for this frequency-dependent shift, we performed a 2D cross-correlation analysis using optically thin regions of the jet in each frequency pair \citep[see also][]{Paraschos21}. We performed the cross-correlation using images produced with natural weighting to increase sensitivity for
the optically thin regions in the extended jet.
For each pair of frequencies, the images were restored with a circular beam whose size corresponds to the arithmetic mean of the equivalent circular beams at the two frequencies. The pixel size was chosen to be one tenth of the beam’s full width at half maximum (FWHM). In this source we found the spectral index distribution in the nuclear region, especially at the highest frequencies, to be very sensitive to the precision of the alignment. Therefore, we adopted a smaller pixel size and the mean beam in order to preserve sufficient resolution in the cross-correlation. Each map was first slightly shifted so that the pixel with peak flux density was centered at the origin, introducing an inherent uncertainty corresponding to one pixel in both axes. After cross-correlation, we estimate the error in the derived shifts as the quadrature sum of the uncertainty of one pixel from the initial alignment and the one-pixel uncertainty of the correlation itself, following the method by \cite{Boccardi2021}. The results of the cross-correlation, including the correlation coefficients and uncertainties, are presented in Table~\ref{tab:table 2}. 

After applying the derived core shifts, we recalculated the radial distances of all \texttt{MODELFIT} components using as the reference origin the most compact region of the two-sided jet base observed at 43 GHz, identified as the midpoint between the jet and counter-jet cores at that frequency. 
All images were then shifted to this common reference frame.

The selection of the reference position is further supported by the spectral index distribution along the jet ridge line (Fig.~\ref{fig:spectral_index}), which shows a maximum of $\alpha \approx +3.2$ near the same location, with $S(\nu) \propto \nu^{\alpha}$. This is consistent with the expectation for a synchrotron self-absorbed core region, likely marking  the position of the supermassive black hole. We restrict our discussion here to this validation aspect of the spectral index profile, while a more detailed analysis of its spatial variation will be presented in (Sect.~\ref{sec:spectral_index}).

\subsection{Stacked images analysis}
\label{sec:stacked images}
\begin{figure*}[!h]
    \centering
        \includegraphics[width=0.49\textwidth]{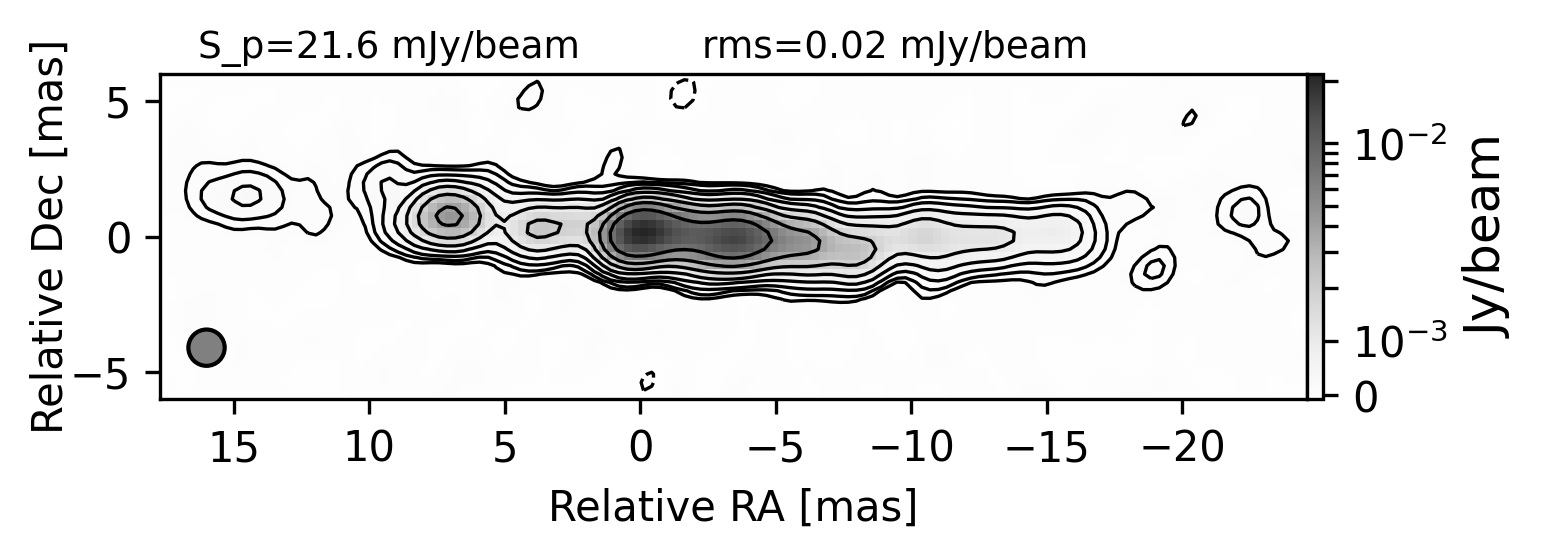}
        \includegraphics[width=0.49\textwidth]{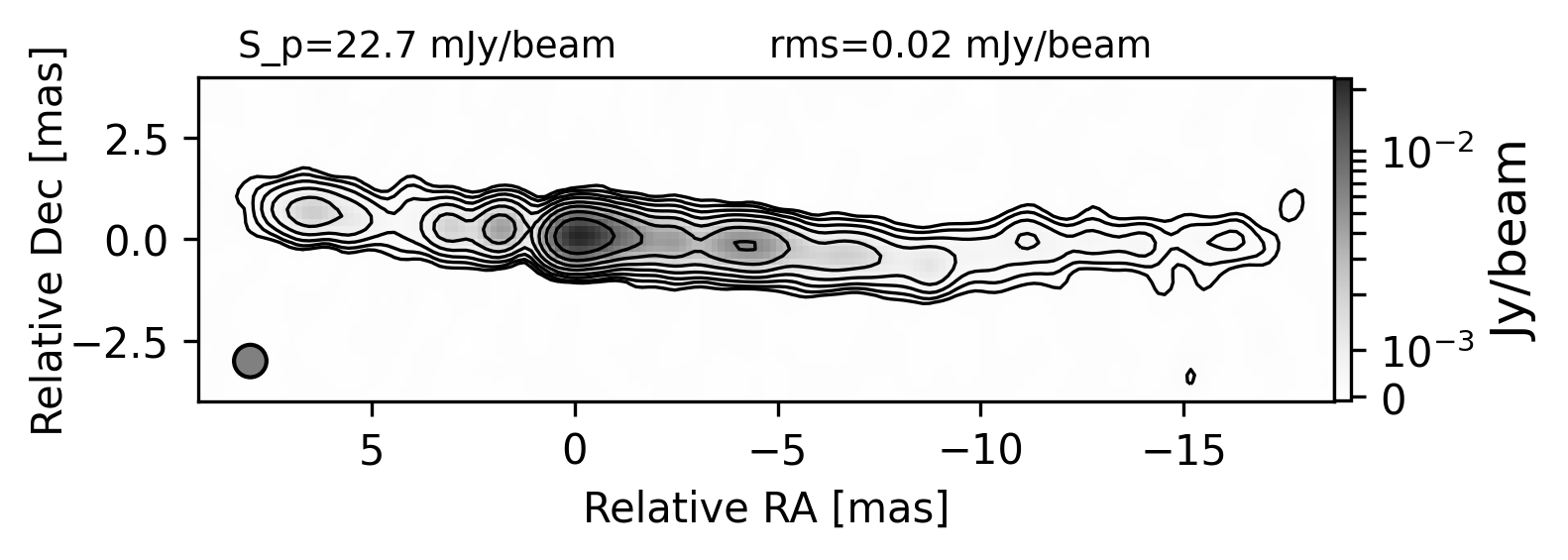}
        \includegraphics[width=0.49\textwidth]{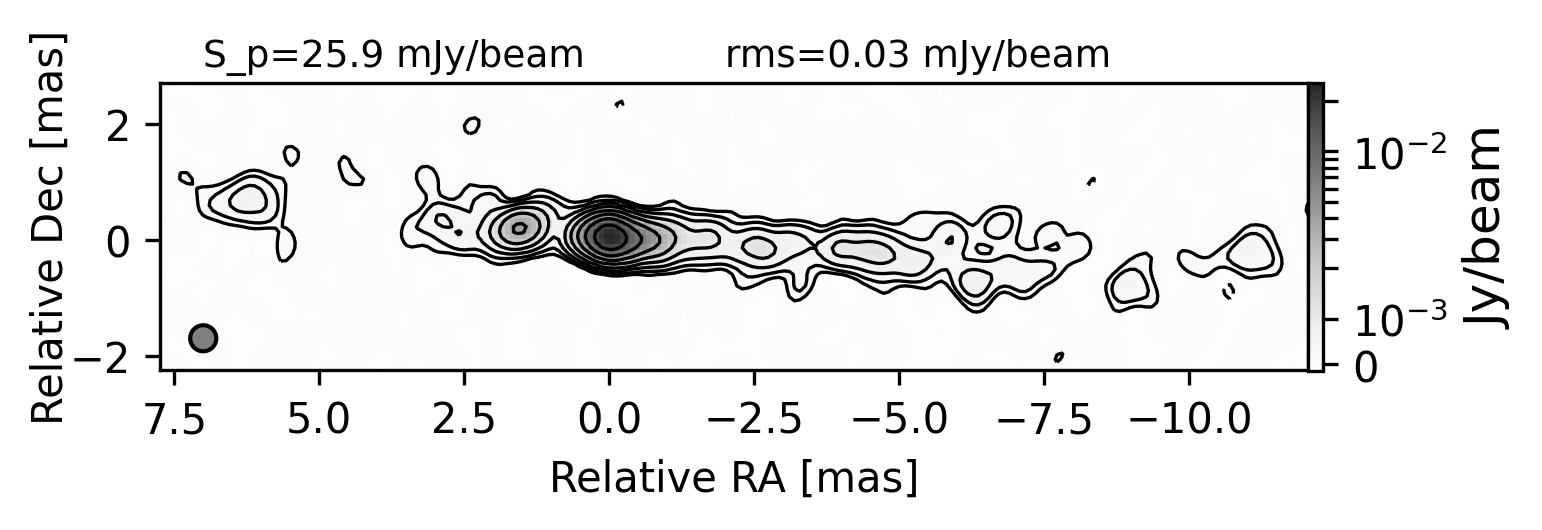}
        \includegraphics[width=0.49\textwidth]{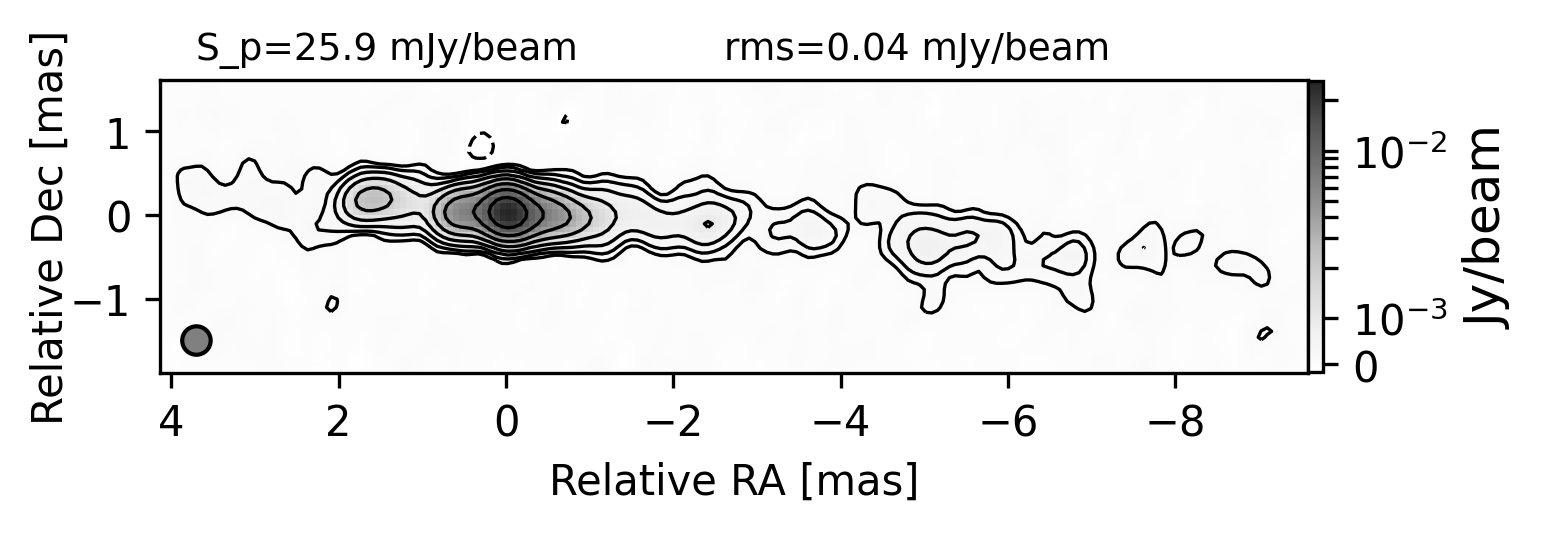}
        \includegraphics[width=0.49\textwidth]{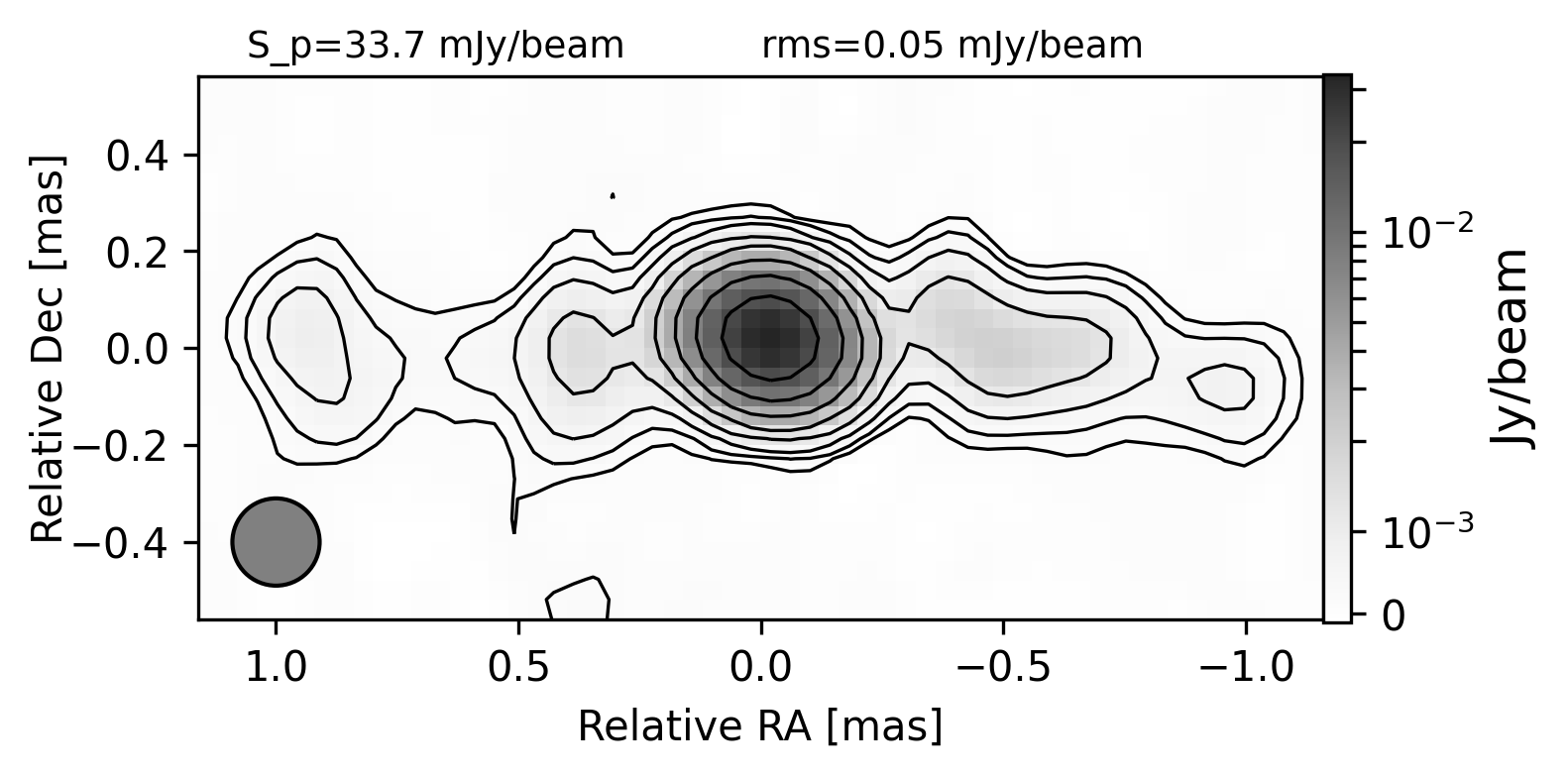}
       \caption{Stacked VLBI images of 3C 452 at 4.9 (upper left), 8.4 (upper right), 15.4 (middle left), 23.6 (middle right), and 43.2 (bottom) GHz. Restored with circular beams of 1.48, 0.8, 0.45, 0.34, and 0.18 mas, respectively. The respective peak flux density and rms values are displayed on the top of each panel. Images were produced using uniform weighting.}
        \label{fig:stacked images}
\end{figure*}

The stacked images at 4.9 GHz, 8.4 GHz, 15.4 GHz, 23.6 GHz, and 43.2 GHz were restored with circular beams of 1.48 mas, 0.8 mas, 0.45 mas, 0.34 mas, and 0.18 mas, respectively. These beam sizes correspond to the average equivalent beams across the observing epochs. The stacked images at 22 GHz and 43 GHz were produced by combining data from our BM516A and BM516B observations with the 2018 observations from project BB393 \citep{2025A&A...695A.118B}. Alignment of the images was performed based on the position of the peak intensity in each epoch. The pixel size for each stacked image was set to $1/5$ of the beam size. We used uniform weighting images for stacking to better define the transverse structure of the jet and allow for precise measurements of width and opening angle. The stacked images are shown in Fig.~\ref{fig:stacked images}.

We performed a pixel-based analysis of the stacked maps using the Python script presented by \cite{Ricci}. The script requires input maps that are aligned along the y-axis and restored with a circular beam. The analysis divides the jet and counter-jet into one-pixel-wide slices oriented perpendicular to the jet's direction. The flux density profile within each slice is then fitted with a single 1D Gaussian function using the Levenberg-Marquardt algorithm (LevMarLSQFitter from the Astropy library). To minimize background noise contamination at the edges of the jet, pixels with values below $3\sigma_\mathrm{rms}$ are excluded. The fitting process ends when the brightest pixel in the slice has an intensity lower than $5\sigma_\mathrm{rms}$.

The flux density in each slice is calculated by integrating the fitted Gaussian profile. The integrated flux and the full width at half maximum (FWHM) of the Gaussian profiles are recorded and utilized to compute the flux density
profiles, the jet opening angle as functions of the distance from the core, and the collimation profiles.

\section{Results}
\label{sec:results}
\subsection{Flux density profiles}

\begin{figure*}[!h]
    \centering
        \includegraphics[width=1\textwidth]{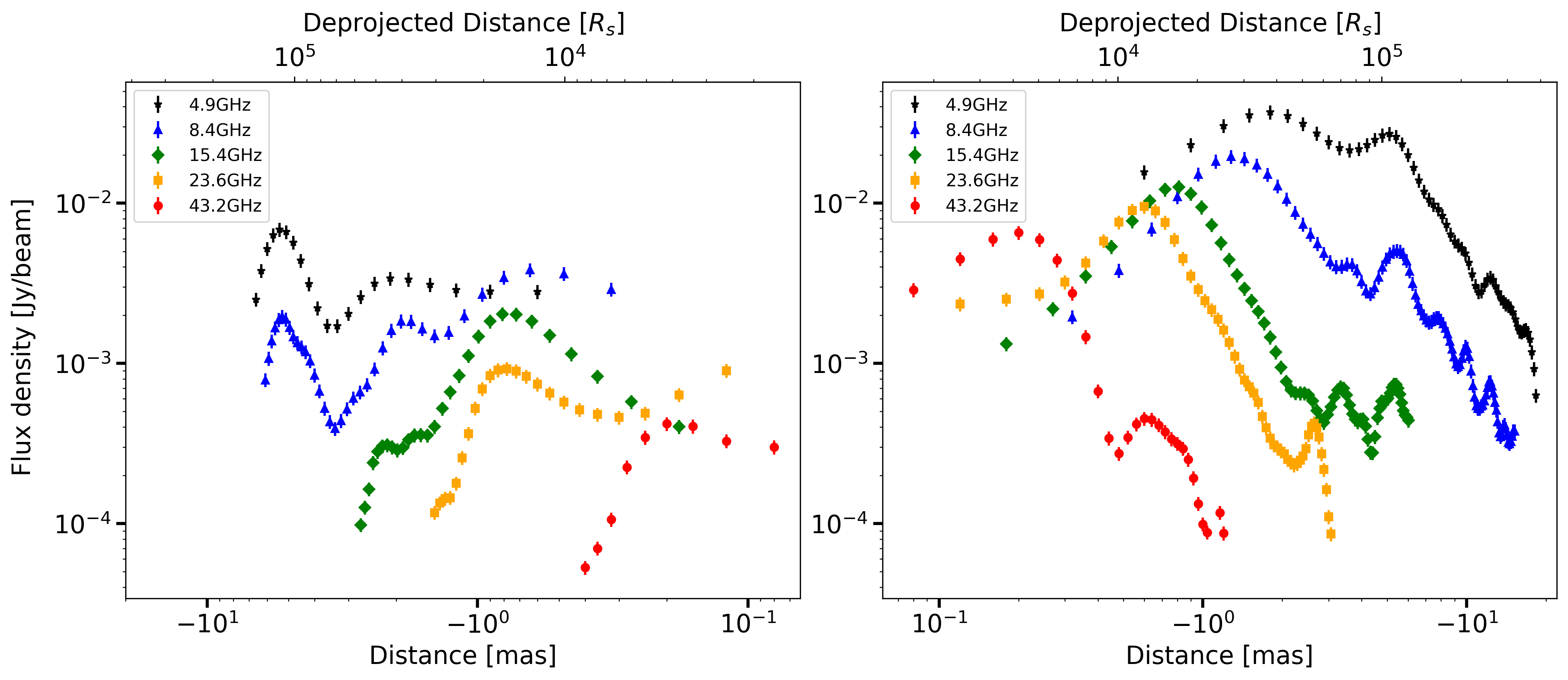}
       \caption{Flux density profiles of the receding counter-jet (left panel) and approaching jet (right panel) and  as a function of distance from the common reference frame, measured from the stacked images at each frequency. The profiles were derived using pixel-based Gaussian fitting across transverse jet slices, as described in Sect.~\ref{sec:stacked images}.}
        \label{fig:stacked images-intensity profiles}
\end{figure*}

The flux density profiles of the approaching jet and receding counter-jet in 3C\,452, derived from the pixel-based analysis of the stacked VLBI images at all observed frequencies, are presented in Fig.~\ref{fig:stacked images-intensity profiles}. Both sides show complex behavior with no clear symmetry in flux density between the jet and counter-jet.

The approaching jet generally displays a decline in flux density with increasing distance from the core, consistent with synchrotron cooling and adiabatic expansion of the relativistic plasma. The decline is more pronounced in the inner few milliarcseconds, particularly at higher frequencies, and becomes less steep further downstream.

In contrast, the counter-jet profiles exhibit a more gradual evolution, with a slower decline compared to the approaching jet. This trend is especially evident at lower frequencies. The systematically lower brightness of the counter-jet across all bands may be primarily due to mild relativistic de-boosting, as expected given the inferred large viewing angle (Sect.~\ref{sec:core temp}). However, absorption effects could also play a role, particularly in the innermost regions. Given the source's classification as a narrow-line radio galaxy and the presence of nuclear obscuration inferred from X-ray observations \citep{fioretti}, both free-free absorption by a circumnuclear torus or dense ionized gas and synchrotron self-absorption in the counter-jet may contribute to the observed brightness asymmetry. In general, while both the jet and counter-jet show flux density gradients, the profiles are not symmetric, differing in both morphology and slope.

\subsection{Jet-to-counter-jet intensity ratio}
\label{sec:jet-to-cjet}

\begin{figure}[!h]  \includegraphics[width=0.49\textwidth] 
{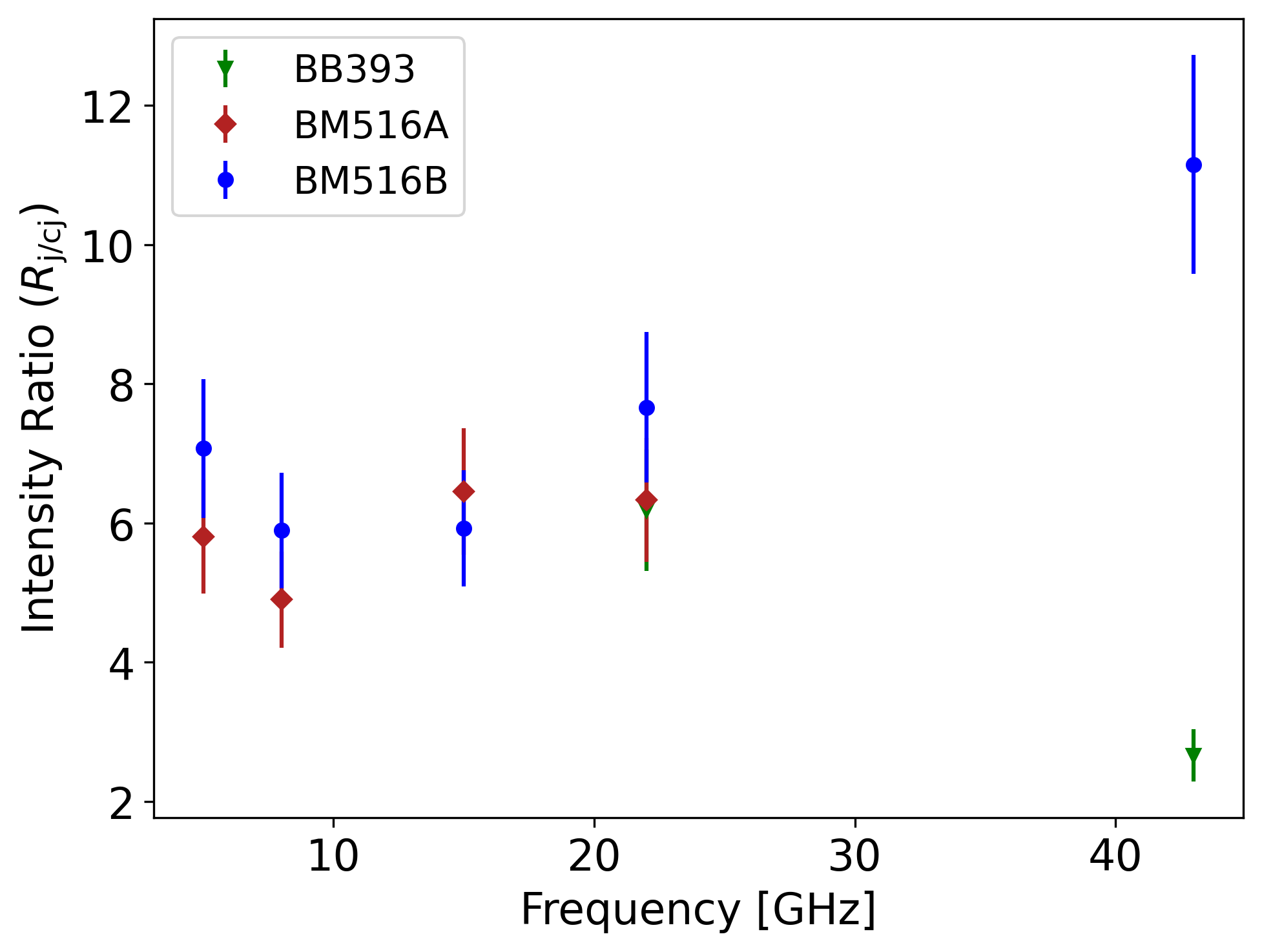}
       \caption{Jet-to-counter-jet intensity ratio as a function of frequency. Ratios are derived for three datasets: BB393, BM516A, and BM516B. All images were aligned to the common reference frame, identified as the midpoint between the jet and counter-jet cores in the 43 GHz map (Sect.~\ref{sec:alignment}). BB393 observations were published in \cite{2025A&A...695A.118B}}
        \label{fig:intensity ratios vs freq}
\end{figure}

\begin{figure}[!h]
        \includegraphics[width=0.49\textwidth]{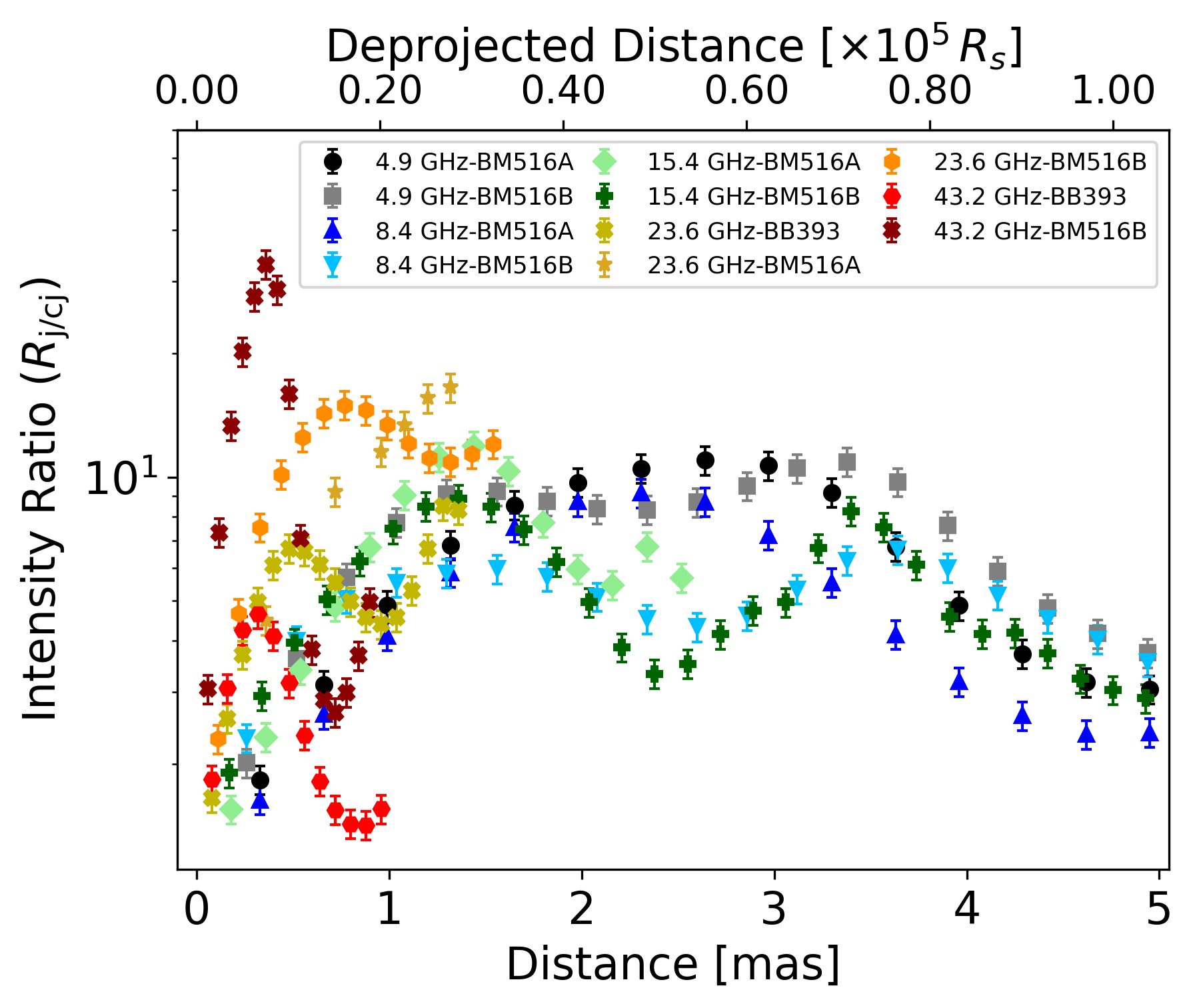}
       \caption{Jet-to-counter-jet intensity ratio as a function of distance from the common reference frame, shown for all observing frequencies. The analysis includes data from projects BB393, BM516A, and BM516B.}
        \label{fig:intensity ratios vs distance}
\end{figure}

The jet-to-counter-jet intensity ratios (\( R_{\mathrm{j}/\mathrm{cj}} \)) were computed after applying the frequency-dependent core shifts listed in Table~\ref{tab:table 2} and aligning all images to the most compact region in the 43.2 GHz map, as discussed in Sect. ~\ref{sec:alignment}.

Fig.~\ref{fig:intensity ratios vs freq} displays the frequency dependence of \( R_{\mathrm{j}/\mathrm{cj}} \) for three observing epochs. Here, \(R_{\mathrm{j}/\mathrm{cj}}\) is derived by summing the  integrate flux densities of the \texttt{MODELFIT} components on each side after shifting the jet and counter-jet components to a common reference frame. The total jet flux density is then divided by the total counter-jet flux density at each frequency. Across most epochs and frequencies, the ratio remains relatively stable at $\sim6.5$.
An exception is observed at 43.2 GHz in the BM516B epoch, where \( R_{\mathrm{j}/\mathrm{cj}} \) increases significantly. To further assess this deviation, we note that the 43.2 GHz core component reported by \cite{2025A&A...695A.118B} has a flux density of \(28.50 \pm 2.27\) mJy when the source was likely in a quiescent state, whereas in our BM516B epoch it is \(42.25 \pm 2.97\) mJy, indicating a real increase in core emission. This increase in core flux density explains the elevated jet-to-counter-jet ratio at this frequency. This outlier is likely related to a transient flaring event in the approaching jet, where newly ejected material temporarily increases the observed ratio. Such events are expected to produce time-dependent brightness asymmetries, as the radiation from the counter-jet side travels a longer path to reach the observer and may not yet reflect the same ejection.

Furthermore, the stability of \( R_{\mathrm{j}/\mathrm{cj}} \) at lower frequencies suggests that, on larger scales, the emission is less affected by nuclear variability and opacity effects, which dominate closer to the core. This allows the jet and counter-jet to appear more comparable in brightness, even though their detailed flux density profiles remain distinct.

To further examine the spatial variation of the intensity, Fig.~\ref{fig:intensity ratios vs distance} plots the jet-to-counter-jet intensity ratio ($R_{\mathrm{j}/\mathrm{cj}}$) as a function of the projected distance along the jet, up to 5 mas ($< 1.05 \times 10^5 \, R_S $). We focus on this scale where the jet acceleration and collimation are expected \citep{Boccardi2021}. A clear trend emerges in the inner 1–1.5 mas ($<3 \times 10^4 \, R_S $), where the ratio exhibits a sharp rise, particularly at higher frequencies (22–43 GHz), reaching values up to $\sim11$. This sharp increase likely reflects an increasing jet speed. Beyond 1.5–2 mas, the ratio fluctuates moderately in the range of 5–8 at lower frequencies (5–15 GHz). This suggests that the jet is moving at roughly constant speed in this region.

\subsection{Core brightness temperature and jet orientation}
\label{sec:core temp}
For a non-thermal source, the apparent brightness temperature can be calculated using the following expression \citep[e.g.][]{kadler}:

\begin{equation}
T_\mathrm{b} = 1.22 \times 10^{12} (1 + z) \left( \frac{S_\nu}{\mathrm{Jy}} \right) \left( \frac{\nu}{\mathrm{GHz}} \right)^{-2} \left( \frac{d}{\mathrm{mas}} \right)^{-2}  \mathrm{K}
\end{equation}

where $S_\nu$ represents the flux density, $\nu$ is the frequency and $d$ denotes the diameter of the emitting region.
From the \texttt{MODELFIT} analysis of our multi-frequency data, we derive the brightness temperature of the core component. 
The uncertainties in the apparent brightness temperature were calculated by propagating the errors associated with the different parameters described in Sect.~\ref{sec:model fitting}. All core component properties and resolution limits used for the $T_{\rm B}$ estimates are reported in Table~\ref{tab:table A3} in the Appendix~\ref{app:appendix}. At 23.6 GHz, in the BM516A dataset, the core size was found to be smaller than the resolution limit. Therefore, we estimated only a lower limit for the brightness temperature by assuming a core size equal to the resolution limit.

In order to estimate the jet Doppler factor, we can make an assumption on the intrinsic brightness temperature value, adopting:
\[
T_{\rm int} = T_{\rm eq} \approx 5 \times 10^{10}\,\mathrm{K}.
\]  
This corresponds to the theoretical value for equipartition between the
particle and the magnetic energy densities \citep{Readhead}, and is  consistent with typical estimates for blazar cores in their quiescent states after correcting for Doppler boosting \citep{Homan}. 

Under this assumption, we compute the Doppler factor from the relation  
\[
T_{\rm obs} = \delta\,T_{\rm int},
\]  
where \(T_{\rm obs}\) is the observed brightness temperature. The results of this analysis are summarized in Table~\ref{tab:table 3}. 

The low values of the Doppler factor ($\delta \sim 0.03$–$0.83$) indicate a significant de-boosting. Alternatively, the low observed brightness temperature may indicate a deviation from equipartition, with $T_{\rm int} < T_{\rm eq} $. In this scenario, we can model the intrinsic temperature as $ T_{\rm int}= \eta ^{1/8.5} T_{\rm eq}$, where $\eta = {u_p}/{u_B} $ represents the ratio of the particle energy density to the magnetic energy density. This relationship suggests that the energy density of the magnetic field exceeds the energy density of the particles ($u_B>u_p$), indicating that the magnetic energy dominates in the center of this source.

To better constrain the jet orientation, we assume that the intrinsic brightness temperature equals the equipartition value, which implies that the low observed core brightness temperatures are due to Doppler de-boosting. Under this assumption, we combine the jet-to-counter-jet ratio ($R_{\mathrm{j}/\mathrm{cj}}$) and the Doppler factor to solve for the viewing angle $\theta$ and the intrinsic speed $\beta$.
The $R_{\mathrm{j}/\mathrm{cj}}$ is defined as:  

\begin{equation}
R_{\mathrm{j}/\mathrm{cj}}= \left( \frac{1 + \beta \cos \theta}{1 - \beta \cos \theta} \right)^{2-\alpha},
\end{equation}

where $\beta$ is the intrinsic jet speed in units of $c$, $\theta$ the viewing angle, and $\alpha$ the spectral index. We computed the ratio $R_{\mathrm{j}/\mathrm{cj}}$ as described in Sect.~\ref{sec:jet-to-cjet}, assuming a spectral index $\alpha = -0.7$ with $S_{\nu} \propto \nu^{\alpha}$. 
Together with the Doppler factor,
\begin{equation}
\delta = \frac{\sqrt{1 - \beta^2}}{1 - \beta \cos \theta}
\end{equation}

these two relations yield estimates of the viewing angle $\theta$ and the intrinsic jet speed $\beta$, reported in Table~\ref{tab:table 3}. We obtain an average $\beta \approx 0.994c$ and
$\theta \approx 70^\circ$, corresponding to a Lorentz factor of
\[
\Gamma = \frac{1}{\sqrt{1 - \beta^2}} \approx 9.1,
\]

under the assumption that the low observed brightness temperatures are primarily caused by Doppler de-boosting.
A lower limit of \(\theta \geq 60^\circ\) was inferred by \citet{giovannini}.  \citet{2025A&A...695A.118B} determine an upper limit of \(\theta_{\rm max} = 78.7^\circ \pm 1.0^\circ\) based on the observed jet-to-counter-jet ratio, and, assuming an intrinsic speed of \(\beta = 0.7c\), they derive \(\theta_{(\beta=0.7)} = 74^\circ\). Although their assumed speed is lower than the values we infer from combining Doppler factor and jet-to-counter-jet ratio, and the jet-to-counter-jet ratio they measure is lower, their derived angle lies near our highest values of \(\theta=73.3^\circ \pm 0.51^\circ\) (Table~\ref{tab:table 3}), indicating general agreement on a large viewing angle.

\begin{table*}[!ht]
    \centering
    \caption{Core brightness temperature of the core component and Doppler factors. }
    \small
    \label{tab:table 3}
    \begin{tabular}{ccccccc}
        \hline
        P.C. & Frequency & $T_\mathrm{core}$  & Doppler factor &  
        \( R_{\mathrm{j}/\mathrm{cj}} \) & $\beta$ & $\theta$\\
        & [GHz] & [10$^{10}$ K] & & & [c]& [deg]\\
        \hline
        BM516A & 4.9  & $0.16\pm0.02$ & $0.032\pm0.004$ & $5.80\pm0.29$   & $0.999$ & $71.67\pm0.50$ \\
               & 8.4  & $0.14\pm0.02$ & $0.028\pm0.004$ & $4.95\pm0.25$ & $0.999$ & $73.27\pm0.51$ \\
               & 15.4 & $0.49\pm0.08$ & $0.098\pm0.016$ & $6.45\pm0.32$ & $0.998$ & $70.56\pm0.50$ \\
               & 23.6 & $>4.17$ & $0.834\pm0.110$ & $6.33\pm0.32$ & $0.831$   & $66.67\pm0.63$ \\
        \hline
        BM516B & 4.9  & $1.50\pm0.25$ & $0.300\pm0.050$ & $7.07\pm0.35$ & $0.981$   & $69.27\pm0.54$ \\
               & 8.4  & $0.37\pm0.05$ & $0.074\pm0.010$ & $5.89\pm0.29$ & $0.999$ & $71.50\pm0.50$ \\
               & 15.4 & $0.95\pm0.11$ & $0.190\pm0.022$ & $5.92\pm0.30$ & $0.992$ & $71.30\pm0.51$ \\
               & 23.6 & $0.24\pm0.03$ & $0.048\pm0.006$ & $7.66\pm0.38$ & $0.999$ & $68.88\pm0.49$ \\
               & 43.2 & $0.30\pm0.04$ & $0.060\pm0.008$ & $11.15\pm0.56$& $0.999$ & $65.21\pm0.48$ \\
        \hline
    \end{tabular}
  \tablefoot{Col. 1: Project code. Col. 2: Frequency. Col. 3: Brightness temperature. Col. 4: Doppler factor. Col. 5: Jet-to-counter-jet ratio. Col. 6: Intrinsic jet speed. Col. 7: Viewing angle.}
\end{table*}

\subsection{Jet shape and opening angle}

\label{sec:jet_shape_opening}

\begin{figure*}[!h]
    \centering
        \includegraphics[width=1\textwidth]{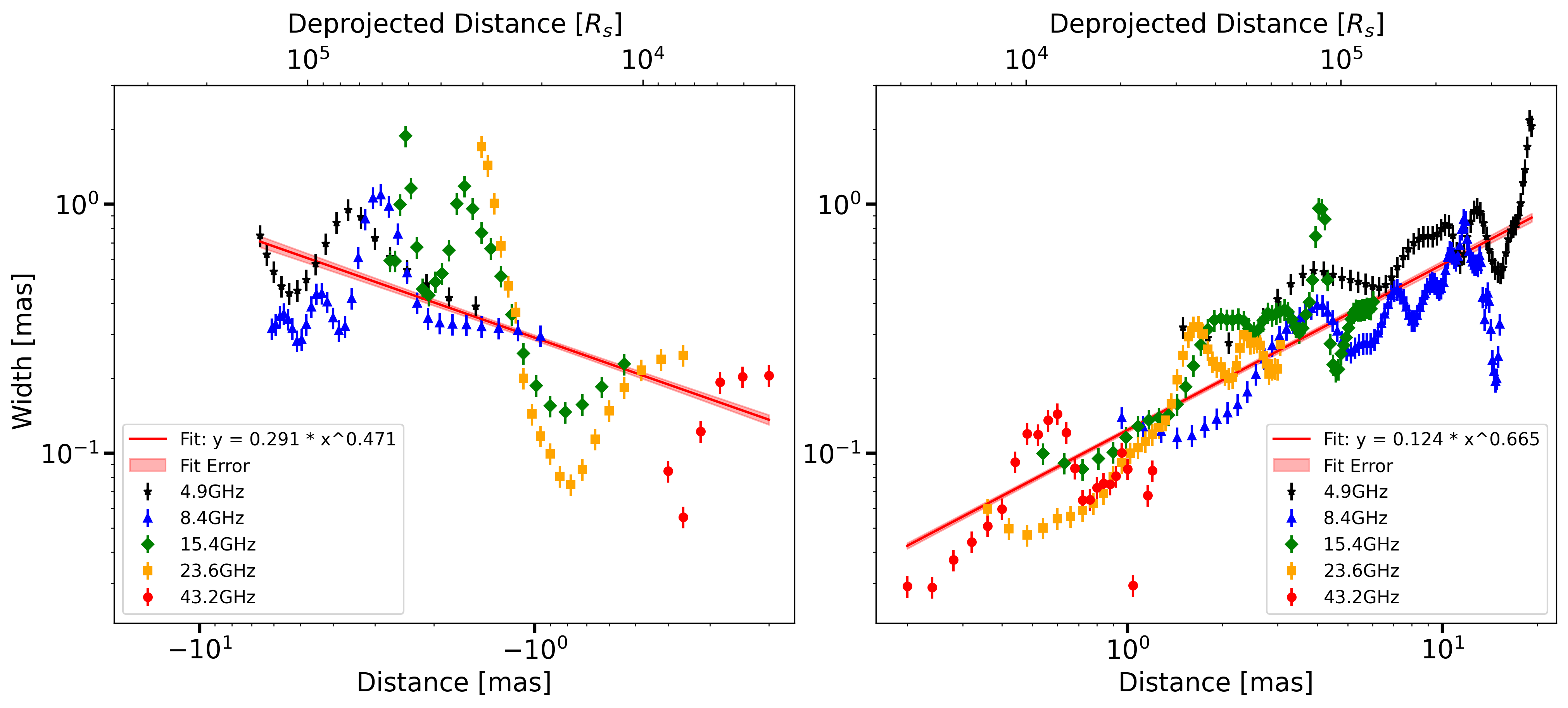}
       \caption{Jet collimation profile of the receding counter-jet (left) and approaching jet (right) derived from stacked images at all frequencies using pixel-by-pixel analysis. The jet width is plotted as a function of distance from the common reference frame. Power-law fits (solid red lines) indicate a parabolic expansion for both sides, with indices $k = 0.47 \pm 0.01$ (counter-jet) and $k = 0.66\pm 0.01$ (jet), consistent with magnetically collimated flows.}
        \label{fig:stacked images-collimation profiles}
\end{figure*}

    The expansion profiles of both the approaching jet and the receding counter-jet were examined, and in Fig.~\ref{fig:stacked images-collimation profiles} we present the two-sided jet width profile of 3C 452, derived from our pixel-by-pixel analysis (Sect.~\ref{sec:stacked images}). In this analysis, we discarded data points located within one beam size from the core at each frequency, for both the jet and counter-jet, to minimize contamination from unresolved core emission.

By fitting a power-law model to describe the jet width $d$ as a function of distance $r$, we found best-fit power-law indices of $k = 0.66 \pm 0.01$ for the jet and $k = 0.47 \pm 0.01$ for the counter-jet, indicating a nearly parabolic shape on both sides.  
This result is further supported by an additional analysis using all \texttt{MODELFIT} components from both epochs (see Appendix~\ref{app:appendix} Fig.~\ref{fig:collimation-modelfit} ), which yielded best-fit power-law indices of $k= 0.554 \pm 0.01$ for the jet and $k = -0.604 \pm 0.01$ for the counter-jet. For this analysis, we excluded components with FWHM smaller than half of the beam minor axis to avoid unresolved features that could bias the fit.
The consistency between these findings and the pixel-by-pixel analysis of the stacked image reinforces the conclusion of a parabolic jet shape.

\begin{figure}
    \centering
        \includegraphics[width=1.0\linewidth]{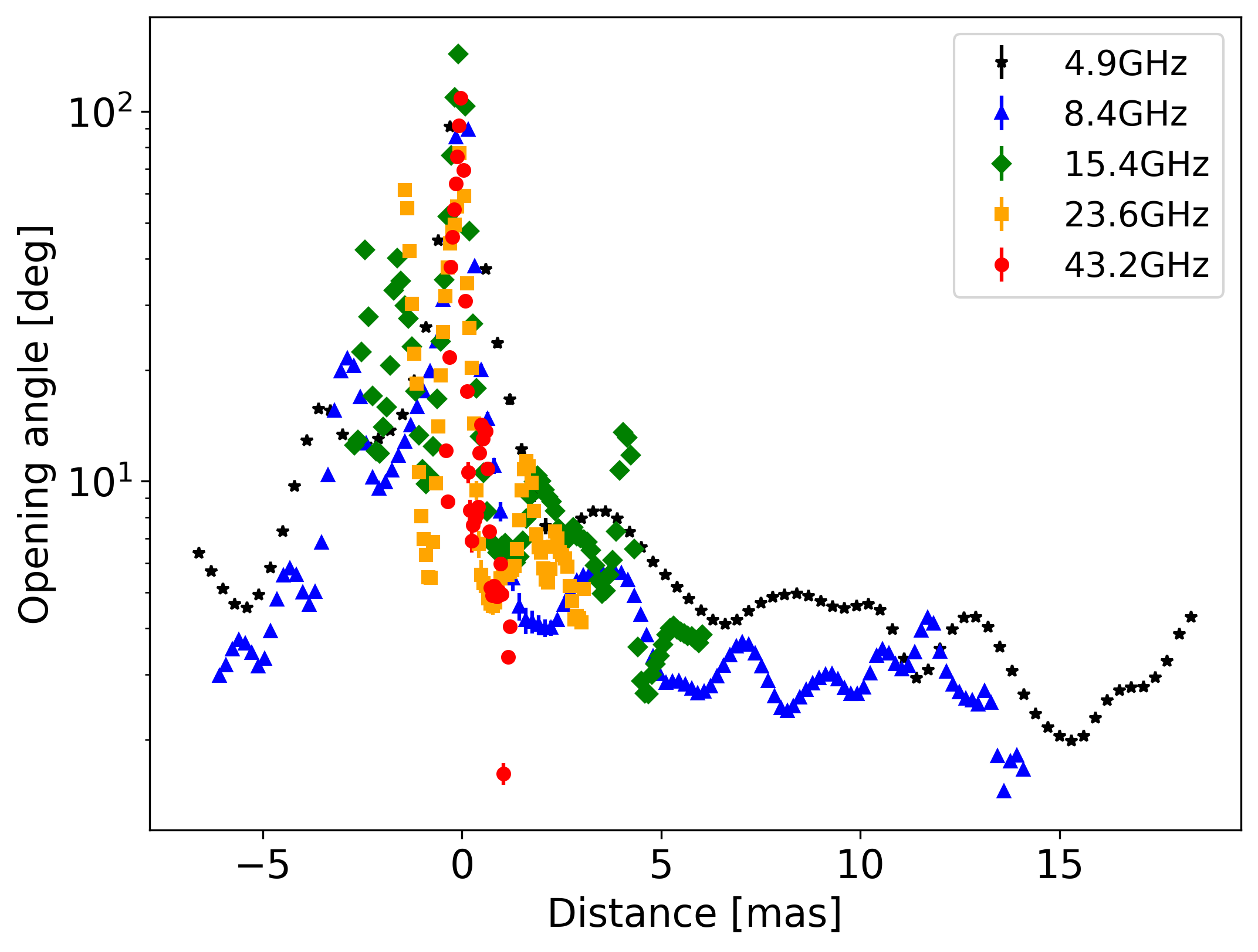}
       \caption{Opening angle profiles of the counter-jet and jet as a function of projected distance from the common reference frame,  derived from pixel-by-pixel Gaussian width fits applied to stacked VLBI images at all frequencies.}
        \label{fig:opening angle}
\end{figure}

To investigate potential changes in jet geometry, we tested for a transition in the jet shape by fitting a broken power law. The half opening angle profile (Fig.~\ref{fig:opening angle}) does indicate the existence of a transition, showing an initial steep decrease and then a flattening with a roughly constant half opening angle beyond 5 mas from the core (corresponding to a de-projected distance of $1.05 \times 10^5 R_{\rm s}$, expressed in Schwarzschild radii). In this region, the half‑opening angle is estimated to be approximately \(4.5^\circ\) (with a full opening angle of ~9°). In our analysis, the apparent half opening angle $\theta_{\rm half}(r)$ is derived from the measured jet width $W(r)$ via the relation $\theta_{\rm half}(r) = \arctan\big[0.5\,W(r)/r\big]$, where $r$ is the projected distance from the core. The observed flattening of the profile beyond 5~mas suggests a possible transition from parabolic to conical geometry in the jet. Parabolic or semi-parabolic shapes are commonly observed in the innermost jet regions (within a few parsecs), with a transition to conical geometry occurring when the collimation ends \citep{kovalev2020, Boccardi2021}.

We modeled the collimation profile using a broken power-law function of the form:
\begin{equation}
d(r) = d_{\rm t} \cdot 2^{(u-w)/h} \cdot \left(\frac{r}{r_{\rm t}}\right)^{u} \cdot \left[1 + \left(\frac{r}{r_{\rm t}}\right)^{h} \right]^{(w-u)/h},
\end{equation}
where $u$ and $w$ are the power-law indices before and after the transition, $d_{\rm t}$ is the jet width at the transition point, $r_{\rm t}$ marks the transition distance and $h$ controls the sharpness of the break. However, the fit did not yield a statistically significant improvement over the single power-law description: the break location was poorly constrained, and the residuals near the transition region remained large. Further analysis with a richer data set providing better sampling of the outer jet regions will be needed for a more robust test, as additional epochs of high-sensitivity, multifrequency observations would increase the signal-to-noise ratio and better recover the faint extended emission in the outer jet, allowing more precise measurements of the jet width on VLBI scales. In addition, complementary lower-frequency and lower-resolution observations would help sample larger spatial scales and provide a more solid description of the jet’s conical region. These complementary data would support a more robust broken power-law fit with tighter constraints on the break point and the power-law indices. Nevertheless, the opening angle profile suggests that a transition is occurring around $10^5 R_{\rm s}$. Interestingly, a significant enhancement in the flux density is observed both in the jet and in the counter-jet side at this distance (Fig.~\ref{fig:stacked images-intensity profiles}). This enhancement is best seen visually in the 5 GHz images (e.g., top left panel in Fig.~\ref{fig:stacked images}) at ~3.5 mas and ~7 mas in the jet and counter jet sides respectively (see also Table~\ref{tab:table A1}). When applying the proper shifts, these features are found to be equidistant from the BH location, and to coincide with the proposed occurrence of the jet shape transition at ~5 mas, or $10^5 R_{\rm s}$. Such bright recollimation features, which are typically stationary, have been observed in other sources at the end of the acceleration and collimation zone \citep[e.g.,][]{boccardi15, Hada2018}, marking the region where the jet stops being magnetically-driven or, alternatively, signaling a  change in the external pressure profile \citep[see also][]{2012ApJ...745L..28A}.
In this context, a comparison with the acceleration estimates derived from the jet-to-counter-jet ratio (Sect.~\ref{sec:jet-to-cjet}) suggests that the bulk of the acceleration occurs within $\lesssim 3 \times 10^{4}\,R_{\mathrm{S}}$, that is, on smaller scales with respect to the putative collimation break near $\sim 10^{5}\,R_{\mathrm{S}}$.

\subsection{Spectral index distribution}
\label{sec:spectral_index}

\begin{figure}
    \centering
    \includegraphics[width=1.0\linewidth]{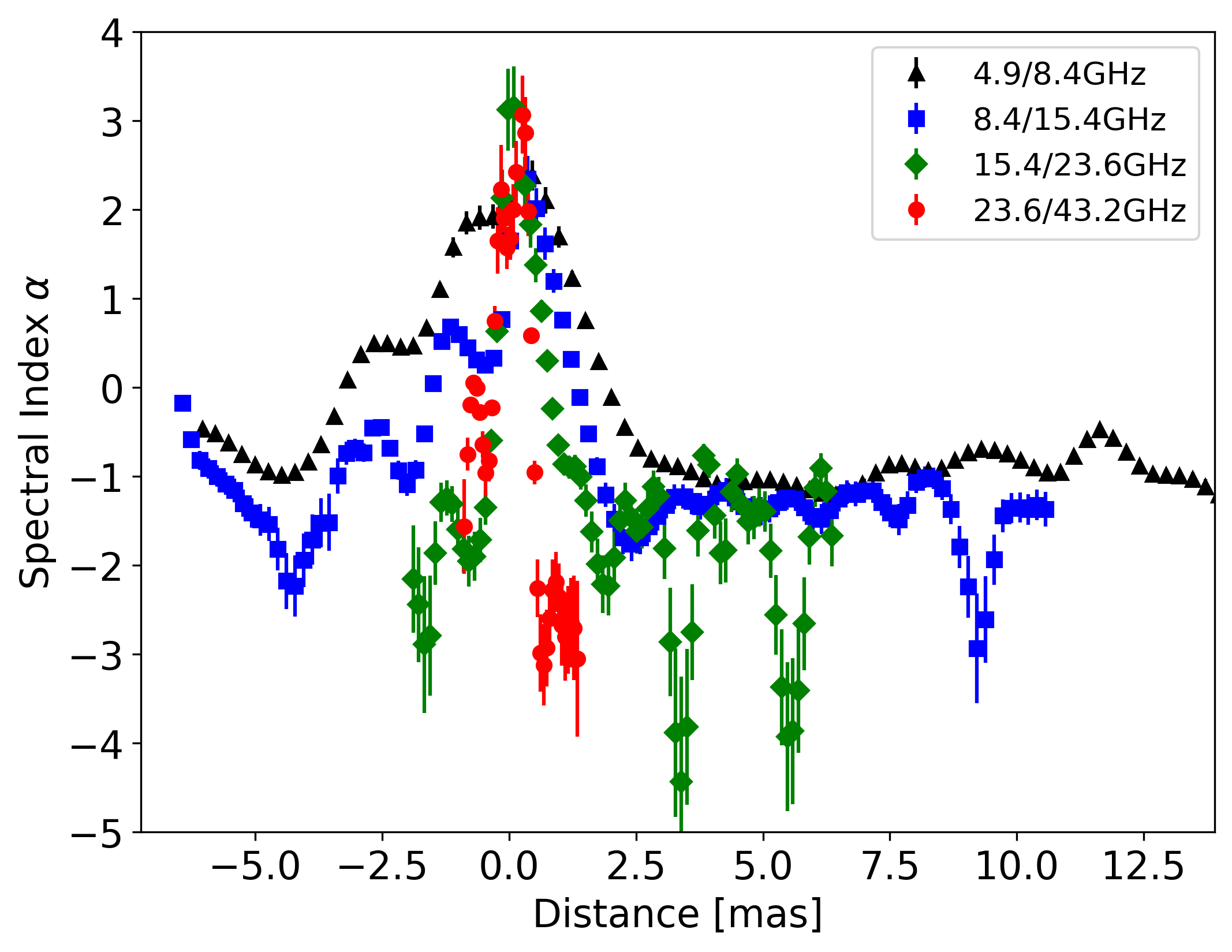}
    \caption{Spectral index profile ($\alpha$) along the jet ridge line of 3C\,452, measured from four frequency pairs: 4.9/8.4 GHz (black triangles), 8.4/15.4 GHz (blue squares), 15.4/23.6 GHz (green diamonds), and 23.6/43.2 GHz (red circles). The horizontal axis shows the projected distance in mas from the common reference frame and the vertical axis shows the spectral index computed via $S_\nu \propto \nu^\alpha$. The central region (within $\sim$2 mas) displays strongly inverted spectra ($\alpha>2$), indicative of synchrotron self-absorption, while downstream values turn negative, consistent with optically-thin synchrotron emission. The peak in spectral index near 0\,mas supports the adopted reference position of the black hole.}
    \label{fig:spectral_index}
\end{figure}

To investigate the spectral properties along the jet, we employed a pixel‑based analysis analogous to that used for our stacked‑image analysis (Sect.~\ref{sec:stacked images}). All natural weighted maps were first aligned in the common reference frame as described in Sect.~\ref{sec:alignment}. For each frequency pair, we convolved both maps to a common circular beam with diameter equal to the average of the two beams and used a pixel size equal to one-fifth of this average beam. 
We divided the jet and counter-jet into one-pixel-wide slices across the jet ridge line. 
We then extracted the flux density in each slice across the jet ridge line by integrating the fitted Gaussian profile. Slices were retained only in regions where the signal in both images exceeded 5 × RMS to ensure reliable measurements.

The spectral index $\alpha$ was calculated for each slice as:

\[
S_{\nu} = S_{0}\,\left(\frac{\nu}{\nu_{0}}\right)^{\alpha},
\]
where $S_{\nu}$ is the flux density at frequency $\nu$, $S_{0}$ is the reference flux at frequency $\nu_{0}$, and $\alpha$ is the spectral index.
 
We included uncertainties by propagating thermal noise (pixel RMS) and calibration errors assumed to be 5\% for the 4.9–8.4 GHz pairs and 10\% for the higher-frequency pairs.

Fig.~\ref{fig:spectral_index} shows $\alpha$ as a function of the projected distance from the common reference frame for four frequency pairs. The core region (within $\sim2\,  $mas) exhibits an inverted spectrum ($\alpha > 2$), indicative of optically thick, synchrotron self-absorbed emission. At the highest frequencies ($>$ 15 GHz) the spectral index exceeds the value of +2.5, which is the theoretical limit for a purely synchrotron self-absorbed emission region. This could indicate that free-free absorption is also taking place, but only in a very compact region at the jet base, which becomes resolved at the highest frequencies. This aspect will be investigated in more detail in a following paper. Further downstream, $\alpha$ becomes steep, reflecting optically thin emission. In the innermost counter-jet and jet region, the spectral index reaches very steep values ($\alpha\sim-2.5$) at the higher frequencies, increasing to more typical optically thin values ($\alpha{\sim}-1$) at larger distances. Similar steep spectral index behavior has been reported in other sources, such as NGC 315 and M 87 \citep{Ricci2025,2023A&A...673A.159R}. In particular, Ricci et al. (2025) suggest that such steep values may result from particles being accelerated via diffusive shock acceleration in strongly magnetized plasma and subsequently undergoing significant synchrotron cooling losses. The recurrence of this behavior across different sources indicates that steep spectral indices may be a common feature of jets in their parabolic collimation region.

\section{Comparison with Cygnus A and other HEGs}
\label{sec:comparison}
The twin-jet systems in 3C\,452 and Cygnus~A offer a unique opportunity to investigate jet collimation and acceleration processes in narrow-line FRII radio galaxies on sub-parsec scales. While Cygnus~A has long served as a benchmark for studies of collimated jet formation in powerful sources, the new multi-frequency VLBI observations of 3C\,452 provide a valuable opportunity to test similar physical processes in a different environment. Both sources exhibit two-sided jets observed at large viewing angles, and are still sufficiently bright on sub-parsec scales at millimeter wavelengths. 

The jet expansion profiles in the two objects indicate a parabolic jet shape, well described by power-law fits. In 3C 452, the jet and counter-jet follow \( r \propto z^{0.66} \) and \( r \propto z^{0.47} \), respectively, with the parabolic geometry persisting to approximately \( 10^5 \, R_S \). In Cygnus~A, the jet shows a similar parabolic profile with \( r \propto z^{0.55} \) up to \( \sim10^4 \, R_S \), beyond which it briefly becomes cylindrical before transitioning to a conical profile \citep{boccardi15}. Therefore, while the jet expansion rates are similar, the extent of the parabolic region is larger in 3C\,452 than in Cygnus~A by an order of magnitude.
Additionally, 3C 452 has an Eddington ratio of \(L/L_{\rm Edd}\approx0.03\) (\(\log(L/L_{\rm Edd})\approx-1.46\) with $M_\mathrm{BH} = {\sim}8 \times 10^8 M_\odot$), based on the uniform hard X-ray selected AGN measurements in the BASS DR2 catalog \citep{Koss2022}. Similarly, Cygnus A has \(L/L_{\rm Edd}\approx0.02\) (\(\log(L/L_{\rm Edd})\approx-1.74\) with $M_\mathrm{BH} = {\sim}2.7 \times 10^9 M_\odot$) from the same catalog. Although Cygnus A hosts a more massive black hole, the similarity in their low  Eddington ratios suggests that the order-of-magnitude difference in the extent of their parabolic jet regions is not driven by gross differences in accretion rate, but instead reflects intrinsic differences in jet collimation and interaction with the surrounding medium.

It has been proposed that jet collimation transitions may occur near the Bondi radius in some AGN, which is the radius at which the black hole’s gravity dominates the thermal motion of the hot gas. In some sources, such as M\,87  and NGC 6251, the parabolic–to–conical jet transition occurs near the Bondi radius, suggesting a possible connection between the external pressure profile and jet collimation \citep[e.g.,][]{Asada2012, Tseng2016}. In other objects (e.g., NGC\,315, NGC\,4261), the transition occurs well inside the Bondi scale \citep[e.g.,][]{Boccardi2021,Balmaverde2008,kovalev2020}. 
The powerful FR\,II Cygnus A maintains a parabolic profile out to $\sim10^4\,R_{\rm S}$ \citep{boccardi15}, well inside the estimated Bondi radius of $\sim3\times10^5\,R_{\rm S}$ \citep{Nakahara2019}. Detailed analysis of the jet width profile in Cygnus A also shows a discontinuity in the jet width profile at distances on the order of $2.3\times10^5$–$6.8\times10^5\,R_{\rm S}$, where the downstream jet appears to widen relative to the upstream trend, roughly at distances near the Bondi radius \citep{Nakahara2019}. For 3C\,452, we cannot make a similar comparison since existing X-ray observations do not provide measurements of the hot gas temperature on scales close to the black hole that are required to estimate the Bondi radius, precluding a meaningful estimate for this source.

It is also instructive to compare the terminal jet opening angles among these powerful radio galaxies. \cite{boccardi15} report an intrinsic full opening angle of approximately \(\sim\!10^\circ\) for Cygnus A. In 3C 452, the apparent full opening angle flattens near \(\sim\!9^\circ\) (Fig.~\ref{fig:opening angle}), which, considering a viewing angle of \(\sim\!70^\circ\), corresponds to an intrinsic full opening angle of \(\sim\!8.4^\circ\), in close agreement with the case of Cygnus A. Such opening angles are large compared to the intrinsic values typically derived for blazars \citep[$\sim1^\circ$, e.g.,][]{2017MNRAS.468.4992P}, and are indicative of mildly relativistic Lorentz factors ($\Gamma\sim1-2$).
Assuming a transversely stratified jet, the Lorentz factor $\Gamma \sim 9$ derived in Sect. 3.3 possibly reflects the bulk speed of the de-boosted filaments of the flow, while the geometrical properties of the radio emission are dominated by a slower component that is the brightest at large viewing angles. At such angles, relativistic Doppler de-boosting suppresses emission from the fast spine, making the slower sheath more prominent and leading to a lower apparent speed. In this case, the high Lorentz factor is obtained by assuming that the low observed brightness temperatures are mainly due to Doppler de-boosting. However, if the low brightness temperatures instead arise from intrinsic jet properties, such as intrinsic high magnetization in an accelerating and magnetically dominated flow, strong Doppler de-boosting would not be required. In that case, similar viewing angles could be obtained with Doppler factors closer to unity, implying lower intrinsic jet speeds and Lorentz factors, consistent with previous studies. Therefore, the derived Lorentz factor depends on the assumed origin of the observed brightness temperatures and does not uniquely determine the intrinsic bulk flow speed.

A notable distinction is that Cygnus A clearly exhibits limb-brightened emission and transversely stratified velocity fields, suggestive of a spine-sheath configuration with a fast central spine surrounded by a slower layer. 
Limb-brightening is not observed in 3C\,452. This may indicate that such transverse structure is intrinsically not present, or alternatively that it remains unresolved. Owing to the higher redshift of 3C 452, the transverse spatial resolution achieved by our 43 GHz VLBI observations corresponds to physical scales of order $\gtrsim10^3 R_{\rm S}$, significantly larger than the $\sim400 R_{\rm S}$ scales resolved in Cygnus A at the same observing frequency \citep{boccardi15}. Therefore, achieving comparable transverse spatial resolution in 3C 452 would require higher-frequency VLBI observations at 86 GHz, which would probe spatial scales of the order of $\lesssim10^3 R_{\rm S}$.

It is also interesting to compare the findings for 3C\,452 and Cygnus\,A with those obtained in the literature for other HEGs. Indeed, jet collimation studies have considered several other objects belonging to this class, but characterized by a smaller viewing angle of the jet. Analysis of the jet expansion profiles in powerful broad-line AGN have revealed very extended collimation zones,
with transitions typically observed beyond \( 10^6 \, R_S \)  (e.g., \cite{Boccardi2021}). In fact, Cygnus A was the only HEG in the sample of \citet{Boccardi2021} with a transition distance consistent with the shorter scales commonly seen in LEGs. Our new results place 3C 452 firmly alongside Cygnus A, making it a second example of an FRII HEG with a comparatively compact collimation region.

\begin{figure}
    \centering
    \includegraphics[width=1.0\linewidth]{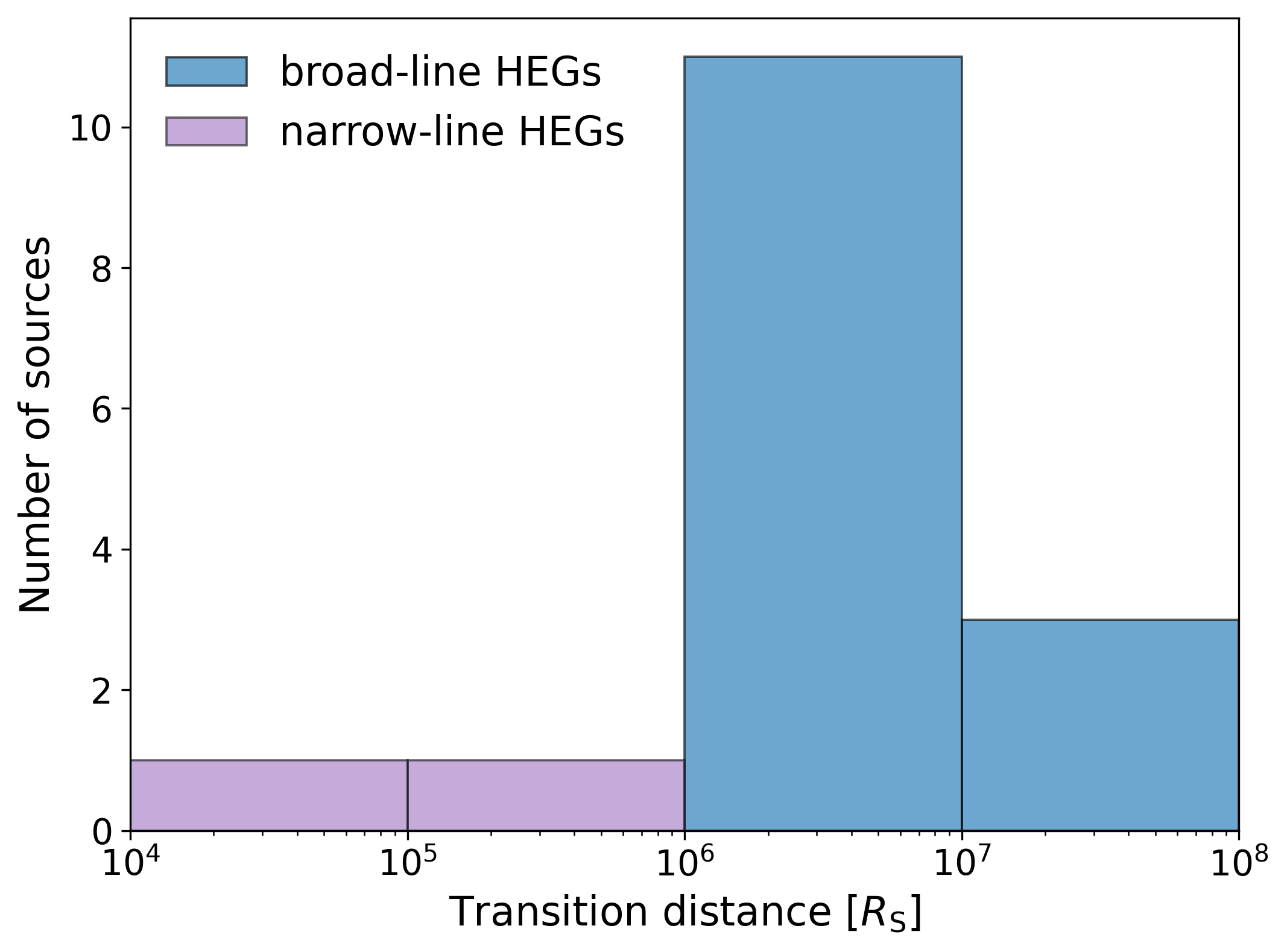}
    \caption{Distribution of the de-projected transition distances in high-excitation galaxies (HEGs), expressed in units of Schwarzschild radii (\(R_\mathrm{S}\)). Data compiled from \citet{Boccardi2021,Algaba2019,Traianou2020,Okino2022,Burd2022,Kunwoo2024,Shang2025}, together with 3C 452. Narrow-line HEGs (Cygnus A and 3C 452) are plotted in purple, while broad-line HEGs are shown in blue.}
    \label{fig:histogram}
\end{figure}

Fig.~\ref{fig:histogram} presents the distribution of transition distances for 15 HEGs reported in the literature \citep{Boccardi2021,Algaba2019,Traianou2020,Okino2022,Burd2022,Kunwoo2024,Shang2025}, together with 3C 452. 
We divided the sample into two groups: narrow-line HEGs viewed at large angles (Cygnus A and 3C 452) and broad-line HEGs (including several FSRQs, such as CTA 26, PKS 0528+134, 4C +71.07, 4C +29.45, and 3C 279) viewed at smaller angles. All broad-line HEGs exhibit jet shape transitions at \( 10^6-10 ^7 \, R_S \). In contrast, Cygnus A and 3C\,452 lie 1–2 orders of magnitude below this range. Since we expect all these sources have intrinsically similar properties, this comparison might indicate that the determination of the jet transition distance has a dependence on the jet orientation. For instance, if the jet does have a spine-sheath transverse structure, it is possible that these two components collimate on different scales.

In sources seen close to the line of sight, such as FSRQs, the spine dominates and appears to collimate out to very large distances, whereas in narrow-line radio galaxies, such as 3C 452 and Cygnus A, the sheath may define the observed break scale.

The number statistics for narrow-line HEGs is clearly still very limited, which highlights the importance of studies considering fainter sources such as 3C\,452 for understanding jet formation in powerful sources.

\section{Conclusions}
\label{sec:conclusions}
We have presented a detailed multi-frequency VLBI study of the twin-jet system in the FRII radio galaxy 3C 452, based on high-resolution observations at 5–43 GHz. The availability of both jet and counter-jet across all frequencies enabled us to probe the jet structure symmetrically on sub-parsec scales, offering rare insights into the collimation and physical conditions at the jet base of an FRII system viewed at a large inclination. Moreover, we note that the double-double morphology of this source, with faint large-scale relic lobes beyond the classical FRII lobes \citep[]{Sirothia}, suggests a restarted activity, while the misalignment between sub-parsec and kiloparsec jet position angles (Fig.~\ref{fig:big-pic}) could indicate variations in nuclear properties over time. Our main results can be summarized as follows:

\begin{itemize}

\item 
We find that the jet-to-counter-jet intensity ratio $R_{\mathrm{j}/\mathrm{cj}}$ remains nearly constant at lower frequencies, implying that the emission is dominated by intrinsic jet properties and is less affected by nuclear variability or opacity effects. 
In contrast, at higher frequencies and within the inner $\sim$1–1.5~mas, $R_{\mathrm{j}/\mathrm{cj}}$ increases sharply, consistent with an increasing jet speed near the jet base. Beyond $\sim2$ mas, the ratio levels off, indicating that the jet speed is approximately constant on larger scales.

\item 
Combined constraints from $R_{\mathrm{j}/\mathrm{cj}}$ and brightness temperatures yield a viewing angle $\theta \approx 70^\circ$ and intrinsic speed $\beta \approx 0.99c$, consistent with the source's classification as a lobe-dominated FRII galaxy. The low Doppler factors ($\delta \sim 0.03\text{--}0.83$) further support minimal beaming effects.
Our orientation constraints are in good agreement with previous estimates, including the lower limit of $\theta \geq 60^\circ$ from \citet{giovannini} and the upper limit of $\theta_{\rm max} = 78.7^\circ \pm 1.0^\circ$ from \citet{2025A&A...695A.118B}.

\item 
Our analysis reveals a remarkable symmetry between the two jets in the collimation properties. The jet and counter-jet follow a near-parabolic expansion, with power-law indices of \( k = 0.66 \pm 0.01 \) and \( k = 0.47 \pm 0.01 \), respectively. 
The opening-angle profile flattens at $\sim10^{5} R_{\mathrm{S}}$, indicating a possible transition in the jet shape. At the same distance, bright symmetric features are observed on both sides, consistent with stationary recollimation structures that may mark the end of the collimation region.
 A comparison with the profile of $R_{\mathrm{j}/\mathrm{cj}}$ with distance indicates that most of the acceleration occurs upstream of this distance.

\item
Our pixel-based spectral index analysis reveals a strongly inverted core spectrum ($\alpha > +2$ within $\sim$2 mas). At the highest frequencies, the core spectrum exceeds the synchrotron self-absorption limit (+2.5), pointing to the presence of additional absorption processes in a very compact region at the jet base. Further out, the spectrum steepens to optically thin values, with very steep indices ($\alpha\sim-2.5$) in the inner jet regions, a behavior also observed in other nearby radio galaxies such as NGC 315 and M 87.

\item 
The distribution of transition distances in a sample of HEGs shows that broad-line sources exhibit jet-shape transitions at $10^{6}$–$10^{7} R_{\mathrm{S}}$,  whereas the narrow-line objects 3C 452 and Cygnus A fall 1-2 orders of magnitude below this range. This suggests that orientation plays an important role in the observed collimation scale.
  \end{itemize}

Taken together, these results show that 3C 452 emerges as a rare and valuable analogue to Cygnus A, allowing us to probe jet collimation and symmetry in a second narrow-line FRII galaxy, and highlighting the importance of orientation in shaping the observed properties of powerful jets in HEGs.

 \begin{acknowledgements}
The authors thank the anonymous referee for their valuable feedback.  
E.M., B.B., and L.R. acknowledge financial support from an Otto Hahn Research Group funded by the Max Planck Society.  
E.M. is a member of the International Max Planck Research School (IMPRS) for Astronomy and Astrophysics at the Universities of Bonn and Cologne.  
This work is based on observations obtained with the Very Long Baseline Array (VLBA) and the Effelsberg 100-m radio telescope.  
The VLBA is a facility of the National Science Foundation operated under cooperative agreement by Associated Universities, Inc.  
The European VLBI Network is a joint facility of European, Chinese, South African, and other radio astronomy institutes funded by their national research councils.
\end{acknowledgements}

\bibliographystyle{aa}
\bibliography{reference.bib}

\begin{thebibliography}{44}
\expandafter\ifx\csname natexlab\endcsname\relax\def\natexlab#1{#1}\fi

\bibitem[{{Algaba} {et~al.}(2019){Algaba}, {Rani}, {Lee}, {Kino}, {Park}, \& {Kim}}]{Algaba2019}
{Algaba}, J.~C., {Rani}, B., {Lee}, S.~S., {et~al.} 2019, \apj, 886, 85

\bibitem[{{Asada} \& {Nakamura}(2012{\natexlab{a}})}]{2012ApJ...745L..28A}
{Asada}, K. \& {Nakamura}, M. 2012{\natexlab{a}}, \apjl, 745, L28

\bibitem[{{Asada} \& {Nakamura}(2012{\natexlab{b}})}]{Asada2012}
{Asada}, K. \& {Nakamura}, M. 2012{\natexlab{b}}, \apjl, 745, L28

\bibitem[{{Balmaverde} {et~al.}(2008){Balmaverde}, {Baldi}, \& {Capetti}}]{Balmaverde2008}
{Balmaverde}, B., {Baldi}, R.~D., \& {Capetti}, A. 2008, \aap, 486, 119

\bibitem[{{Black} {et~al.}(1992){Black}, {Baum}, {Leahy}, {Perley}, {Riley}, \& {Scheuer}}]{black}
{Black}, A.~R.~S., {Baum}, S.~A., {Leahy}, J.~P., {et~al.} 1992, \mnras, 256, 186

\bibitem[{{Blandford} {et~al.}(2019){Blandford}, {Meier}, \& {Readhead}}]{Blandford2019}
{Blandford}, R., {Meier}, D., \& {Readhead}, A. 2019, \araa, 57, 467

\bibitem[{{Blandford} \& {K{\"o}nigl}(1979)}]{Blandford79}
{Blandford}, R.~D. \& {K{\"o}nigl}, A. 1979, \apj, 232, 34

\bibitem[{{Boccardi} {et~al.}(2016){Boccardi}, {Krichbaum}, {Bach}, {Mertens}, {Ros}, {Alef}, \& {Zensus}}]{boccardi15}
{Boccardi}, B., {Krichbaum}, T.~P., {Bach}, U., {et~al.} 2016, \aap, 585, A33

\bibitem[{{Boccardi} {et~al.}(2017){Boccardi}, {Krichbaum}, {Ros}, \& {Zensus}}]{Boccardi2017}
{Boccardi}, B., {Krichbaum}, T.~P., {Ros}, E., \& {Zensus}, J.~A. 2017, \aapr, 25, 4

\bibitem[{{Boccardi} {et~al.}(2021){Boccardi}, {Perucho}, {Casadio}, {Grandi}, {Macconi}, {Torresi}, {Pellegrini}, {Krichbaum}, {Kadler}, {Giovannini}, {Karamanavis}, {Ricci}, {Madika}, {Bach}, {Ros}, {Giroletti}, \& {Zensus}}]{Boccardi2021}
{Boccardi}, B., {Perucho}, M., {Casadio}, C., {et~al.} 2021, \aap, 647, A67

\bibitem[{{Boccardi} {et~al.}(2025){Boccardi}, {Ricci}, {Madika}, {Bartolini}, {Bach}, {Grandi}, {Torresi}, {Krichbaum}, \& {Zensus}}]{2025A&A...695A.118B}
{Boccardi}, B., {Ricci}, L., {Madika}, E., {et~al.} 2025, \aap, 695, A118

\bibitem[{{Burd} {et~al.}(2022){Burd}, {Kadler}, {Mannheim}, {Baczko}, {Ringholz}, \& {Ros}}]{Burd2022}
{Burd}, P.~R., {Kadler}, M., {Mannheim}, K., {et~al.} 2022, \aap, 660, A1

\bibitem[{{Buttiglione} {et~al.}(2010){Buttiglione}, {Capetti}, {Celotti}, {Axon}, {Chiaberge}, {Macchetto}, \& {Sparks}}]{Buttiglione2010}
{Buttiglione}, S., {Capetti}, A., {Celotti}, A., {et~al.} 2010, \aap, 509, A6

\bibitem[{{Fanaroff} \& {Riley}(1974)}]{fanaroff}
{Fanaroff}, B.~L. \& {Riley}, J.~M. 1974, \mnras, 167, 31P

\bibitem[{{Fioretti} {et~al.}(2013){Fioretti}, {Angelini}, {Mushotzky}, {Koss}, \& {Malaguti}}]{fioretti}
{Fioretti}, V., {Angelini}, L., {Mushotzky}, R.~F., {Koss}, M., \& {Malaguti}, G. 2013, \aap, 555, A44

\bibitem[{{Giovannini} {et~al.}(2001){Giovannini}, {Cotton}, {Feretti}, {Lara}, \& {Venturi}}]{giovannini}
{Giovannini}, G., {Cotton}, W.~D., {Feretti}, L., {Lara}, L., \& {Venturi}, T. 2001, \apj, 552, 508

\bibitem[{{Greisen}(1990)}]{Greisen}
{Greisen}, E.~W. 1990, in Acquisition, Processing and Archiving of Astronomical Images, ed. G.~{Longo} \& G.~{Sedmak}, 125--142

\bibitem[{{Hada} {et~al.}(2018){Hada}, {Doi}, {Wajima}, {D'Ammando}, {Orienti}, {Giroletti}, {Giovannini}, {Nakamura}, \& {Asada}}]{Hada2018}
{Hada}, K., {Doi}, A., {Wajima}, K., {et~al.} 2018, \apj, 860, 141

\bibitem[{{Heckman} \& {Best}(2014)}]{HB2014}
{Heckman}, T.~M. \& {Best}, P.~N. 2014, \araa, 52, 589

\bibitem[{{Homan} {et~al.}(2021){Homan}, {Cohen}, {Hovatta}, {Kellermann}, {Kovalev}, {Lister}, {Popkov}, {Pushkarev}, {Ros}, \& {Savolainen}}]{Homan}
{Homan}, D.~C., {Cohen}, M.~H., {Hovatta}, T., {et~al.} 2021, \apj, 923, 67

\bibitem[{{Kadler} {et~al.}(2004){Kadler}, {Ros}, {Lobanov}, {Falcke}, \& {Zensus}}]{kadler}
{Kadler}, M., {Ros}, E., {Lobanov}, A.~P., {Falcke}, H., \& {Zensus}, J.~A. 2004, \aap, 426, 481

\bibitem[{{Kharb} {et~al.}(2010){Kharb}, {Lister}, \& {Cooper}}]{Kharb2010}
{Kharb}, P., {Lister}, M.~L., \& {Cooper}, N.~J. 2010, \apj, 710, 764

\bibitem[{{Koss} {et~al.}(2022){Koss}, {Ricci}, {Trakhtenbrot}, {Oh}, {den Brok}, {Mej{\'\i}a-Restrepo}, {Stern}, {Privon}, {Treister}, {Powell}, {Mushotzky}, {Bauer}, {Ananna}, {Balokovi{\'c}}, {B{\"a}r}, {Becker}, {Bessiere}, {Burtscher}, {Caglar}, {Congiu}, {Evans}, {Harrison}, {Heida}, {Ichikawa}, {Kamraj}, {Lamperti}, {Pacucci}, {Ricci}, {Riffel}, {Rojas}, {Schawinski}, {Temple}, {Urry}, {Veilleux}, \& {Williams}}]{Koss2022}
{Koss}, M.~J., {Ricci}, C., {Trakhtenbrot}, B., {et~al.} 2022, \apjs, 261, 2

\bibitem[{{Kovalev} {et~al.}(2020){Kovalev}, {Pushkarev}, {Nokhrina}, {Plavin}, {Beskin}, {Chernoglazov}, {Lister}, \& {Savolainen}}]{kovalev2020}
{Kovalev}, Y.~Y., {Pushkarev}, A.~B., {Nokhrina}, E.~E., {et~al.} 2020, \mnras, 495, 3576

\bibitem[{{Krause} {et~al.}(2019){Krause}, {Shabala}, {Hardcastle}, {Bicknell}, {B{\"o}hringer}, {Chon}, {Nawaz}, {Sarzi}, \& {Wagner}}]{Krause2019}
{Krause}, M. G.~H., {Shabala}, S.~S., {Hardcastle}, M.~J., {et~al.} 2019, \mnras, 482, 240

\bibitem[{Leahy {et~al.}(1996)Leahy, Bridle, \& Strom}]{Leahy1996_Atlas}
Leahy, J.~P., Bridle, A.~H., \& Strom, R.~G. 1996, {An Atlas of DRAGNs (3CRR Atlas)}, \url{http://www.jb.man.ac.uk/atlas/}, online archive of radio images for 3CRR sources, accessed via NASA NED

\bibitem[{{Lee} {et~al.}(2008){Lee}, {Lobanov}, {Krichbaum}, {Witzel}, {Zensus}, {Bremer}, {Greve}, \& {Grewing}}]{2008AJ....136..159L}
{Lee}, S.-S., {Lobanov}, A.~P., {Krichbaum}, T.~P., {et~al.} 2008, \aj, 136, 159

\bibitem[{{Lobanov}(1998)}]{Lobanov98}
{Lobanov}, A.~P. 1998, \aap, 330, 79

\bibitem[{{Nakahara} {et~al.}(2019){Nakahara}, {Doi}, {Murata}, {Nakamura}, {Hada}, \& {Asada}}]{Nakahara2019}
{Nakahara}, S., {Doi}, A., {Murata}, Y., {et~al.} 2019, \apj, 878, 61

\bibitem[{{Okino} {et~al.}(2022){Okino}, {Akiyama}, {Asada}, {G{\'o}mez}, {Hada}, {Honma}, {Krichbaum}, {Kino}, {Nagai}, {Bach}, {Blackburn}, {Bouman}, {Chael}, {Crew}, {Doeleman}, {Fish}, {Goddi}, {Issaoun}, {Johnson}, {Jorstad}, {Koyama}, {Lonsdale}, {Lu}, {Mart{\'\i}-Vidal}, {Matthews}, {Mizuno}, {Moriyama}, {Nakamura}, {Pu}, {Ros}, {Savolainen}, {Tazaki}, {Wagner}, {Wielgus}, \& {Zensus}}]{Okino2022}
{Okino}, H., {Akiyama}, K., {Asada}, K., {et~al.} 2022, \apj, 940, 65

\bibitem[{{Paraschos} {et~al.}(2021){Paraschos}, {Kim}, {Krichbaum}, \& {Zensus}}]{Paraschos21}
{Paraschos}, G.~F., {Kim}, J.~Y., {Krichbaum}, T.~P., \& {Zensus}, J.~A. 2021, \aap, 650, L18

\bibitem[{{Pushkarev} {et~al.}(2017){Pushkarev}, {Kovalev}, {Lister}, \& {Savolainen}}]{2017MNRAS.468.4992P}
{Pushkarev}, A.~B., {Kovalev}, Y.~Y., {Lister}, M.~L., \& {Savolainen}, T. 2017, \mnras, 468, 4992

\bibitem[{{Readhead}(1994)}]{Readhead}
{Readhead}, A. C.~S. 1994, \apj, 426, 51

\bibitem[{{Ricci} {et~al.}(2022){Ricci}, {Boccardi}, {Nokhrina}, {Perucho}, {MacDonald}, {Mattia}, {Grandi}, {Madika}, {Krichbaum}, \& {Zensus}}]{Ricci}
{Ricci}, L., {Boccardi}, B., {Nokhrina}, E., {et~al.} 2022, \aap, 664, A166

\bibitem[{{Ricci} {et~al.}(2025){Ricci}, {Boccardi}, {R{\"o}der}, {Perucho}, {Mattia}, {Kadler}, {Benke}, {Bartolini}, {Krichbaum}, \& {Madika}}]{Ricci2025}
{Ricci}, L., {Boccardi}, B., {R{\"o}der}, J., {et~al.} 2025, \aap, 693, A172

\bibitem[{{Ro} {et~al.}(2023){Ro}, {Kino}, {Sohn}, {Hada}, {Park}, {Nakamura}, {Cui}, {Yi}, {Chung}, {Hodgson}, {Kawashima}, {An}, {Trippe}, {Algaba}, {Kim}, {Sawada-Satoh}, {Wajima}, {Shen}, {Cheng}, {Cho}, {Jiang}, {Jung}, {Lee}, {Niinuma}, {Oh}, {Tazaki}, {Zhao}, {Akiyama}, {Honma}, {Lee}, {Lu}, {Zhang}, {Asada}, {Cui}, {Hagiwara}, {Hirota}, {Kawaguchi}, {Koyama}, {Lee}, {Oh}, {Sugiyama}, {Takamura}, {Wang}, {Hwang}, {Jung}, {Kim}, {Kim}, {Kobayashi}, {Oh}, {Oyama}, {Roh}, \& {Yeom}}]{2023A&A...673A.159R}
{Ro}, H., {Kino}, M., {Sohn}, B.~W., {et~al.} 2023, \aap, 673, A159

\bibitem[{{Shang} {et~al.}(2025){Shang}, {Zhao}, {Hong}, \& {Hu}}]{Shang2025}
{Shang}, H., {Zhao}, W., {Hong}, X., \& {Hu}, X.-z. 2025, \apj, 986, 198

\bibitem[{{Shelton} {et~al.}(2011){Shelton}, {Hardcastle}, \& {Croston}}]{Shelton}
{Shelton}, D.~L., {Hardcastle}, M.~J., \& {Croston}, J.~H. 2011, \mnras, 418, 811

\bibitem[{{Shepherd}(1997)}]{Shepherd}
{Shepherd}, M.~C. 1997, in Astronomical Society of the Pacific Conference Series, Vol. 125, Astronomical Data Analysis Software and Systems VI, ed. G.~{Hunt} \& H.~{Payne}, 77

\bibitem[{{Sirothia} {et~al.}(2013){Sirothia}, {Gopal-Krishna}, \& {Wiita}}]{Sirothia}
{Sirothia}, S.~K., {Gopal-Krishna}, \& {Wiita}, P.~J. 2013, \apjl, 765, L11

\bibitem[{{Traianou} {et~al.}(2020){Traianou}, {Krichbaum}, {Boccardi}, {Angioni}, {Rani}, {Liu}, {Ros}, {Bach}, {Sokolovsky}, {Lisakov}, {Kiehlmann}, {Gurwell}, \& {Zensus}}]{Traianou2020}
{Traianou}, E., {Krichbaum}, T.~P., {Boccardi}, B., {et~al.} 2020, \aap, 634, A112

\bibitem[{{Tseng} {et~al.}(2016){Tseng}, {Asada}, {Nakamura}, {Pu}, {Algaba}, \& {Lo}}]{Tseng2016}
{Tseng}, C.-Y., {Asada}, K., {Nakamura}, M., {et~al.} 2016, \apj, 833, 288

\bibitem[{{V{\'e}ron-Cetty} \& {V{\'e}ron}(2006)}]{veron}
{V{\'e}ron-Cetty}, M.~P. \& {V{\'e}ron}, P. 2006, \aap, 455, 773

\bibitem[{{Yi} {et~al.}(2024){Yi}, {Park}, {Nakamura}, {Hada}, \& {Trippe}}]{Kunwoo2024}
{Yi}, K., {Park}, J., {Nakamura}, M., {Hada}, K., \& {Trippe}, S. 2024, \aap, 688, A94

\end{thebibliography}

\clearpage

\begin{appendix}

\onecolumn

\section{Images and \texttt{MODELFIT} parameters}
\label{app:appendix}

\begin{figure*}[h!]
    \centering
        \includegraphics[width=0.49\textwidth]{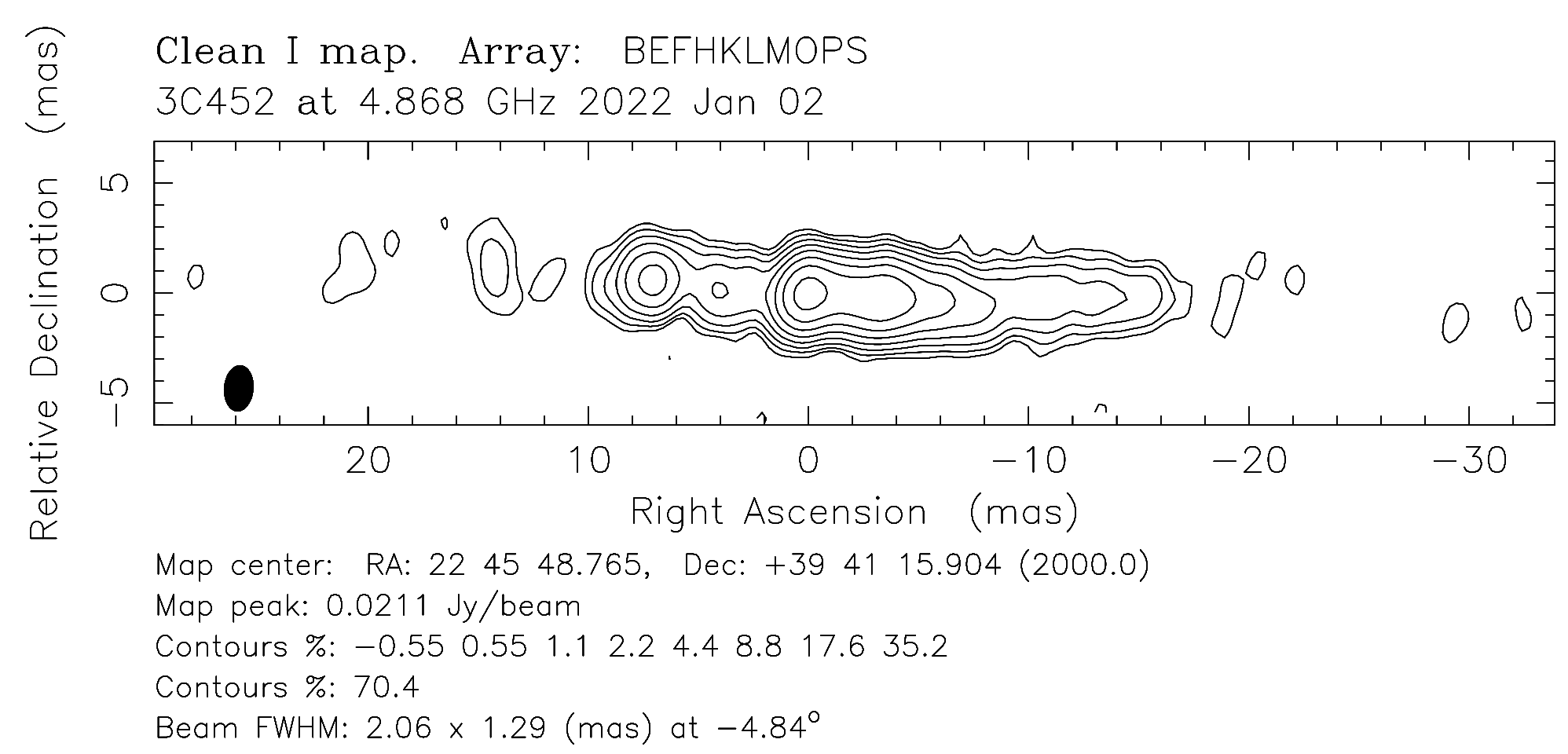}
        \includegraphics[width=0.49\textwidth]{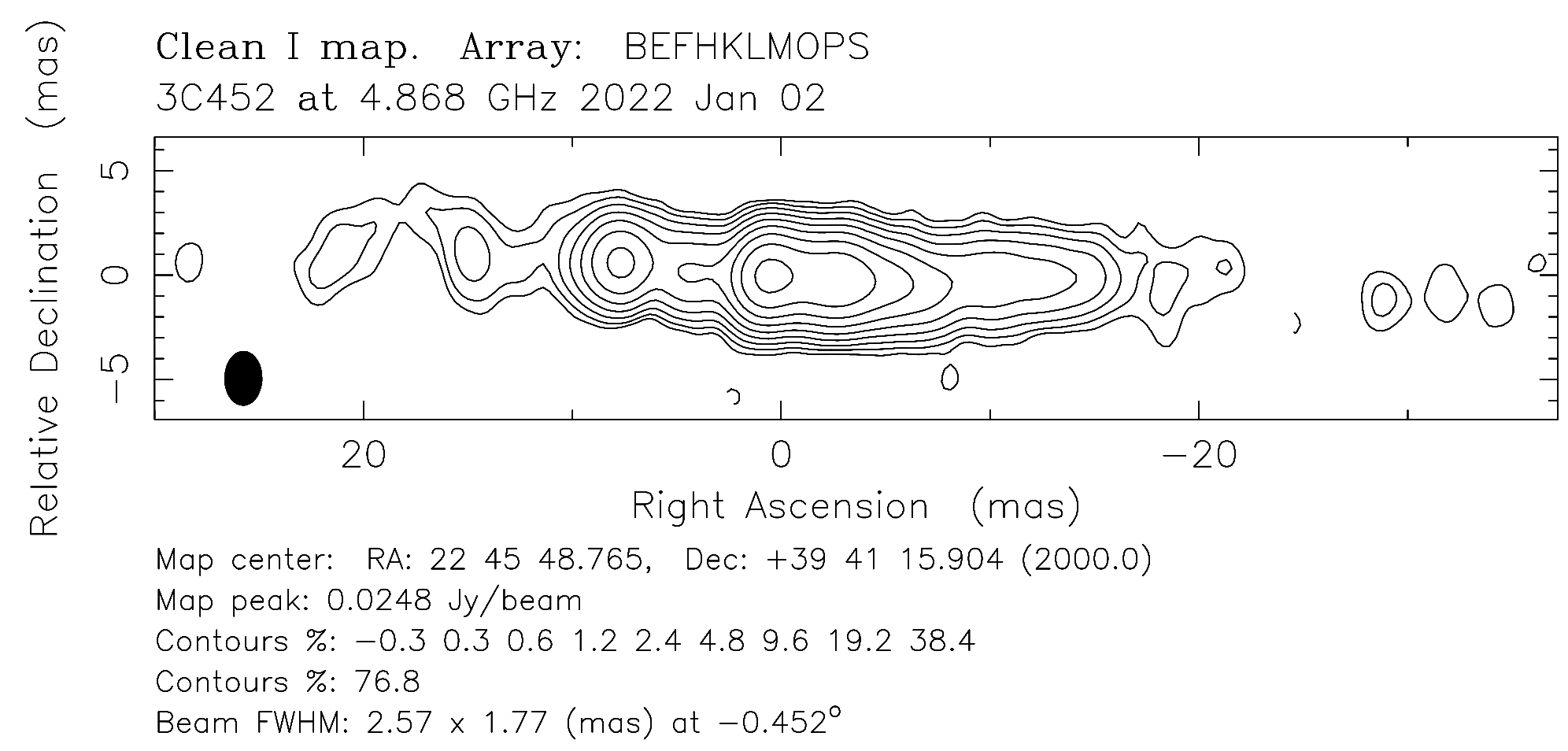}
        \includegraphics[width=0.49\textwidth]{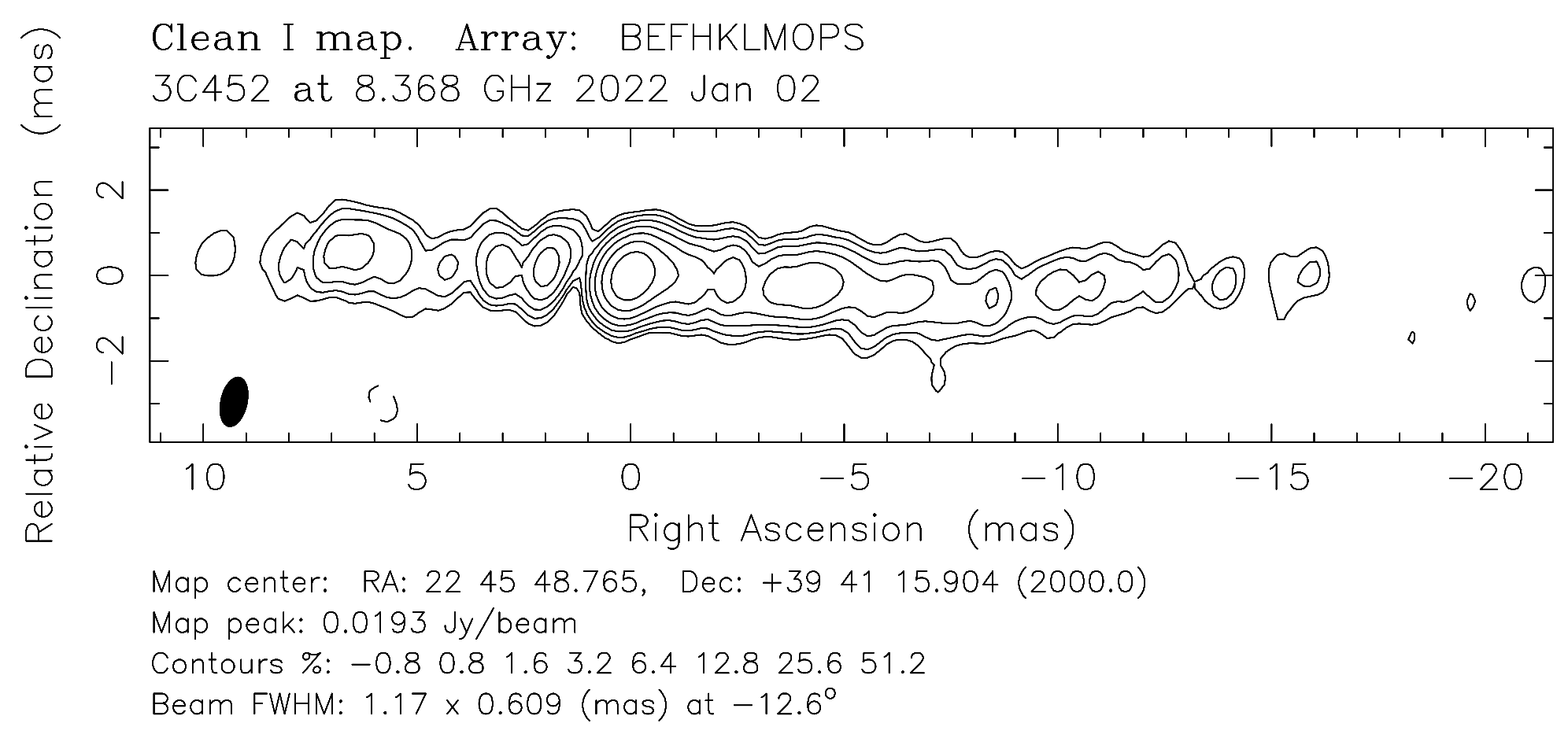}
          \includegraphics[width=0.49\textwidth]{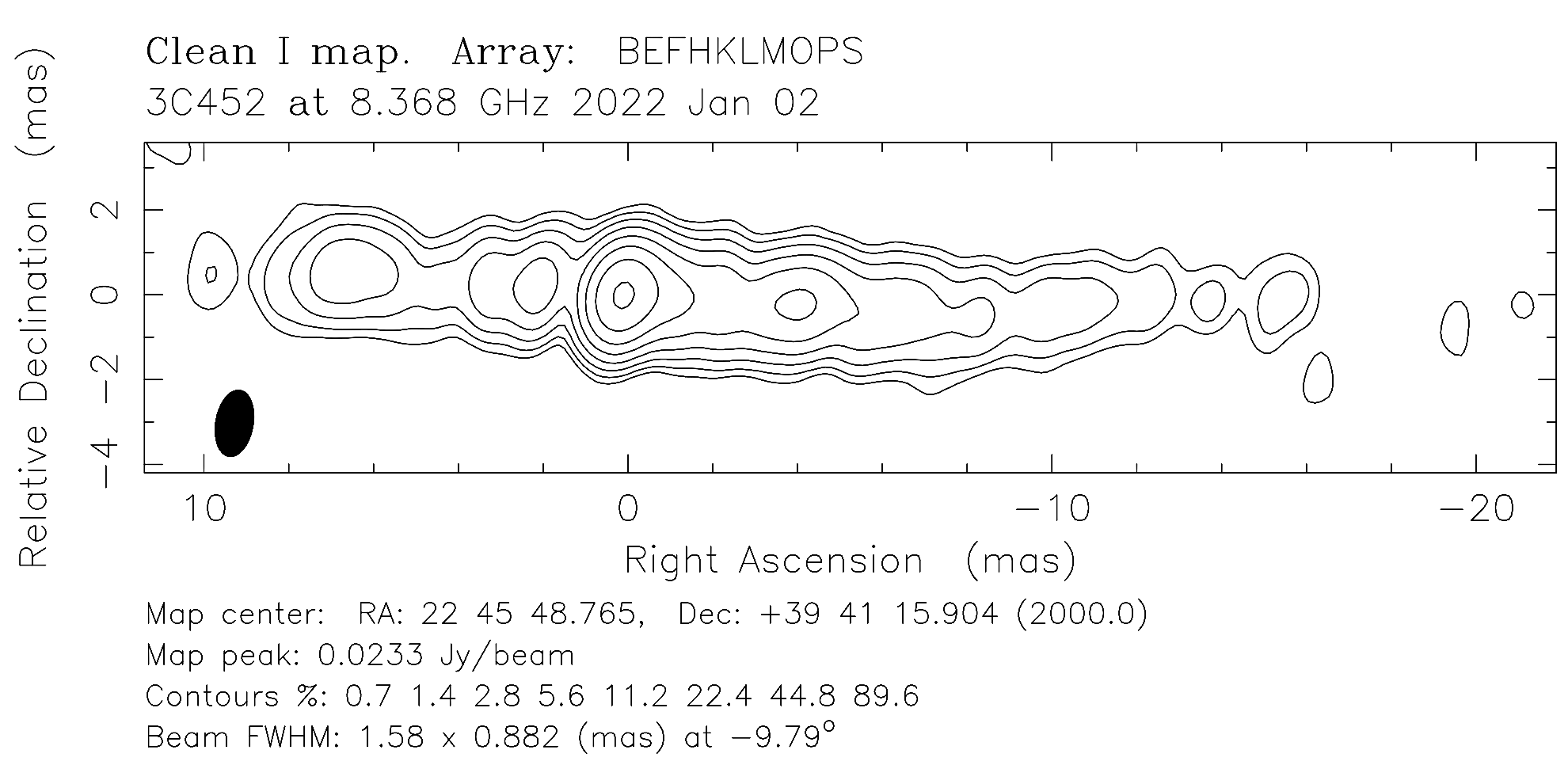}
        \includegraphics[width=0.49\textwidth]{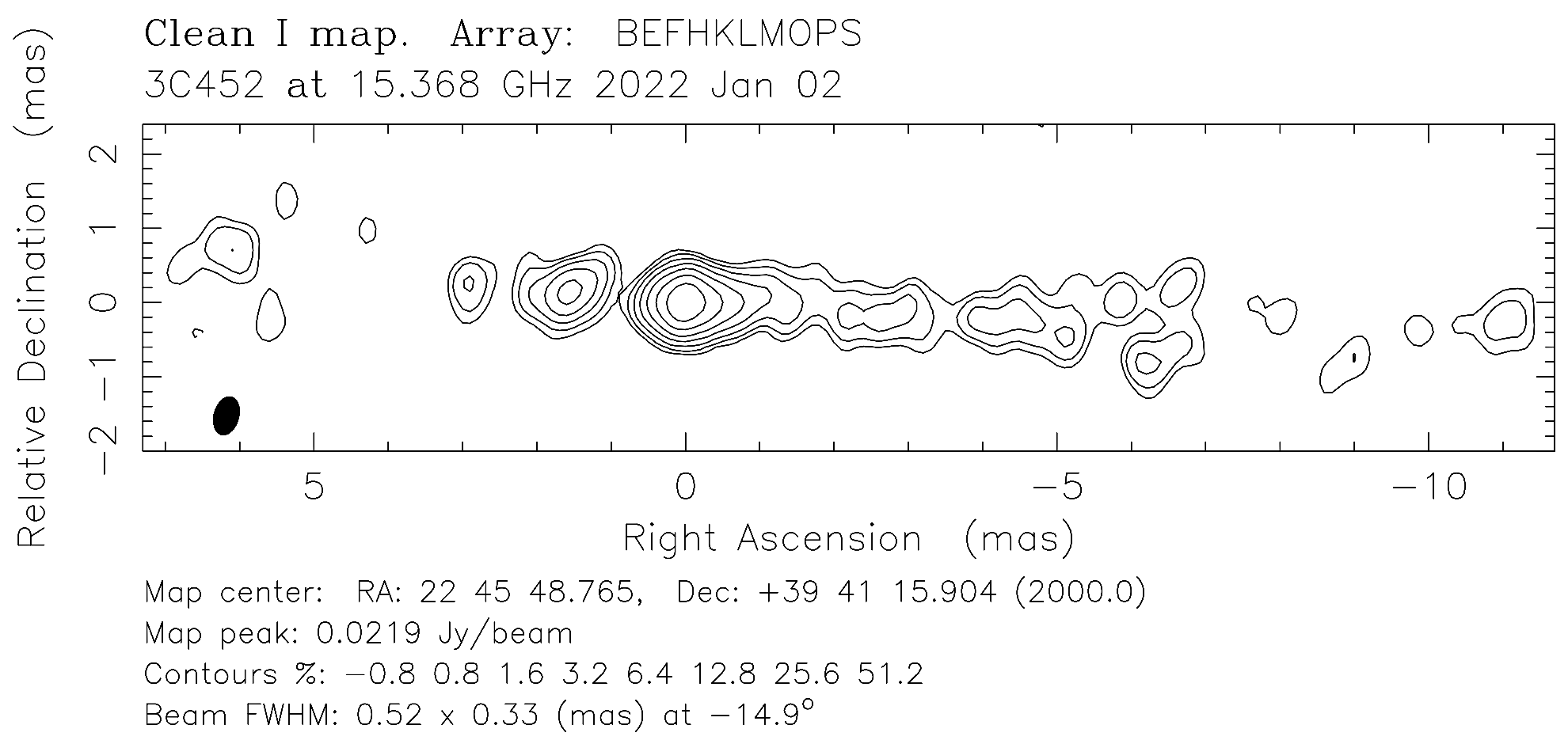}
        \includegraphics[width=0.49\textwidth]{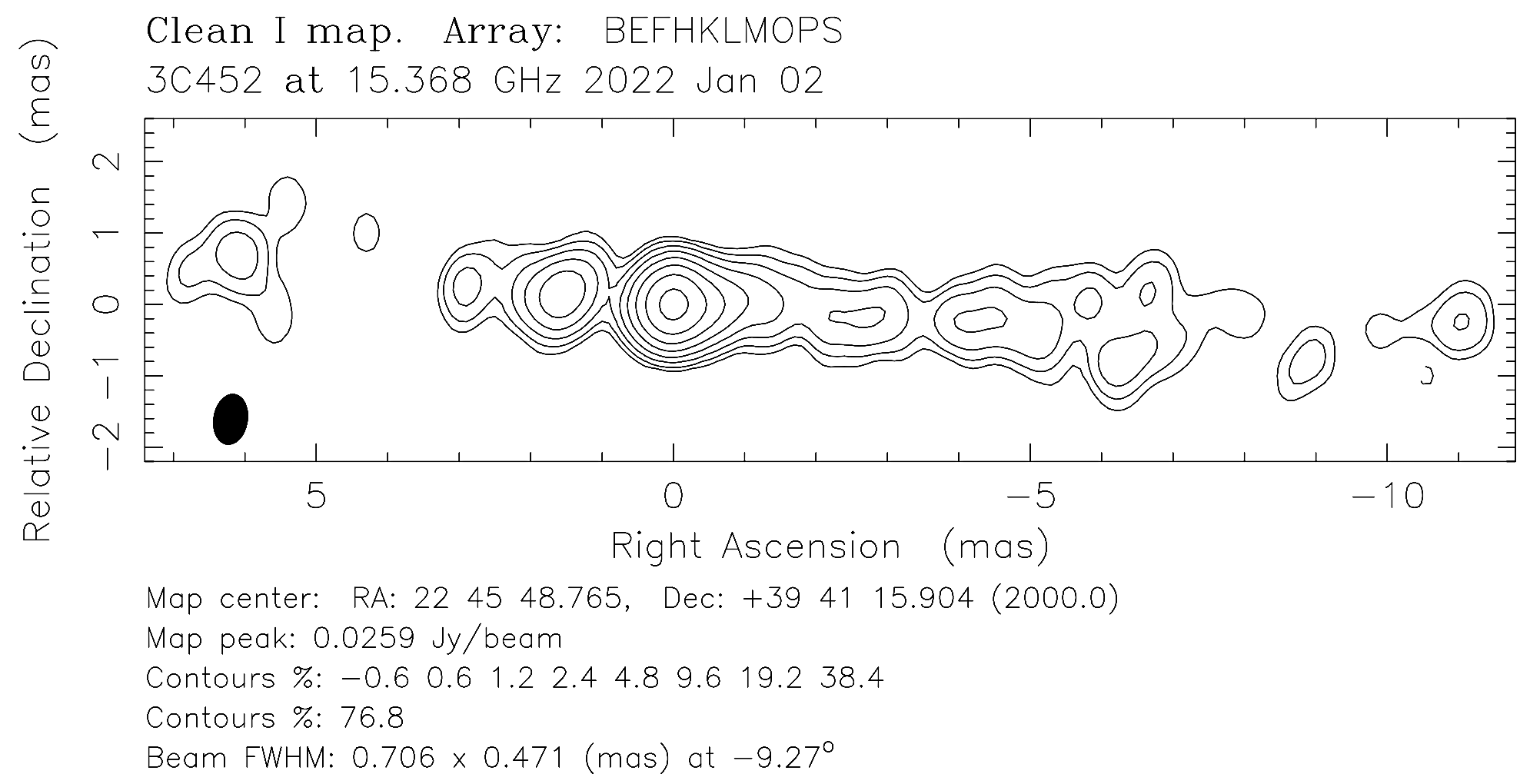}
        \includegraphics[width=0.49\textwidth]{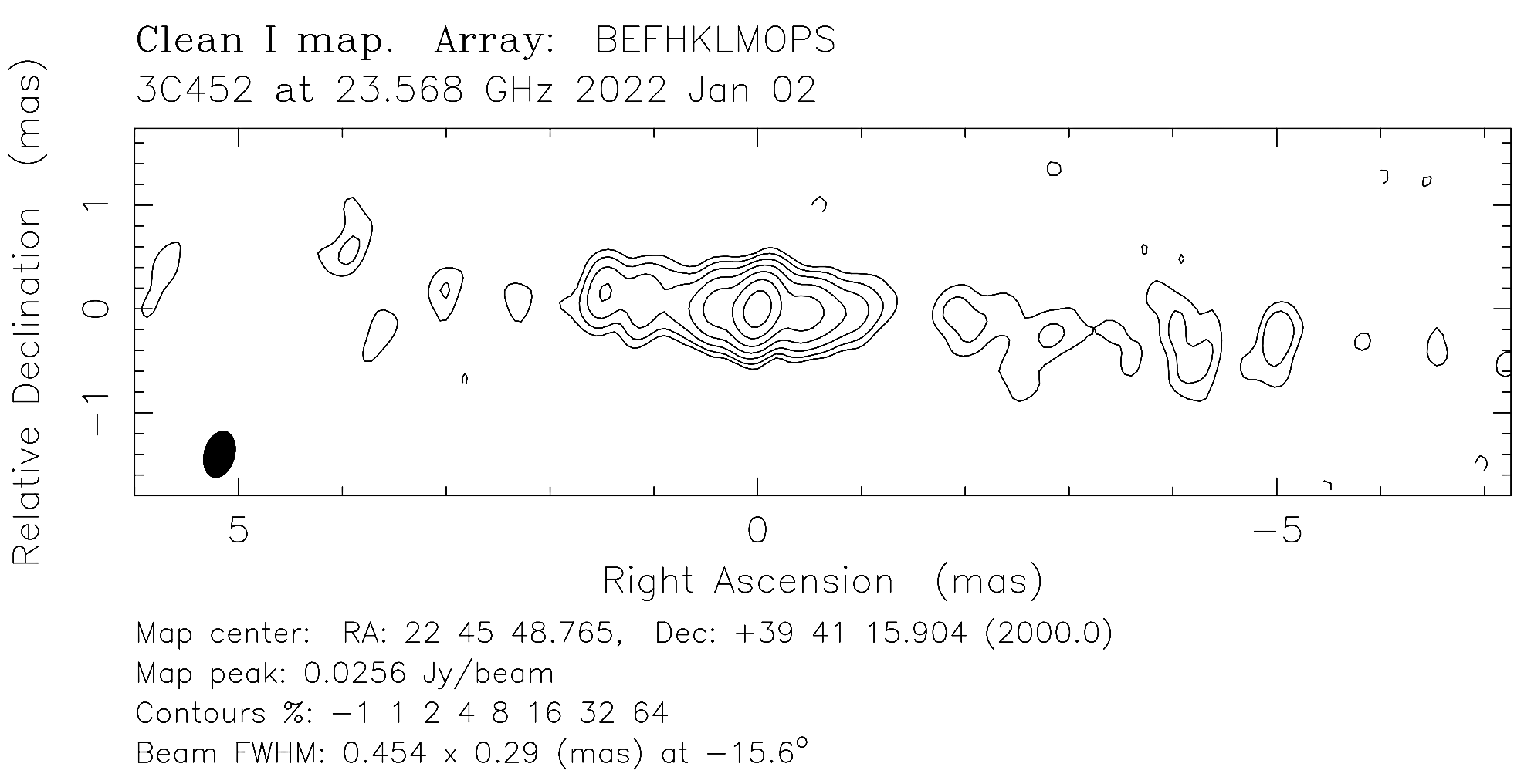}
        \includegraphics[width=0.49\textwidth]{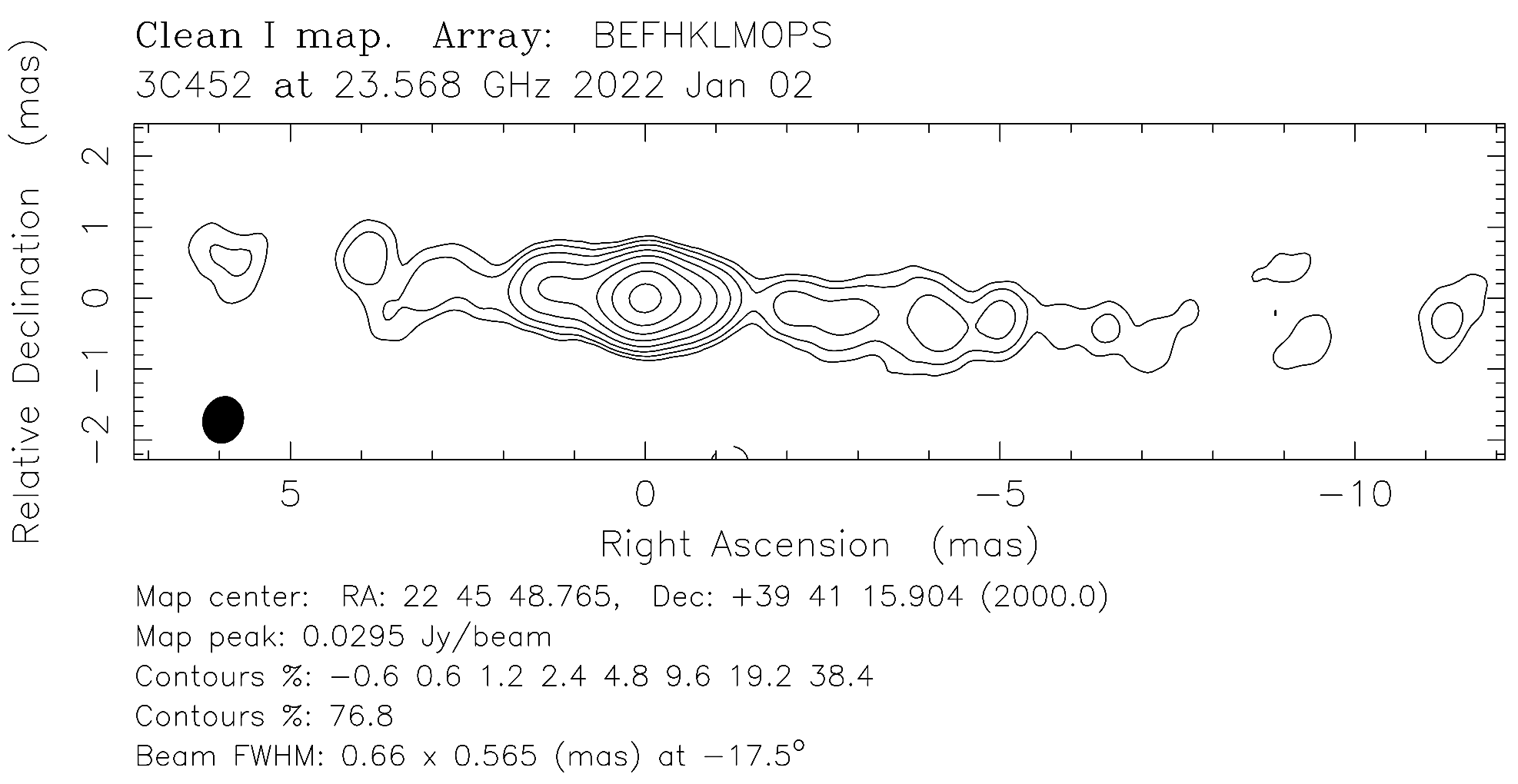}
       \caption{VLBI images of 3C 452 from project BM516A. From top to bottom: 4.9 GHz, 8.4 GHz, 15.4 GHz, and 23.6 GHz. Left panels: images with uniform weighting. Right panels: images with natural weighting.}
    \label{fig:epoch-A-uniform}
\end{figure*}

\begin{figure*}[!h]
    \centering
        \includegraphics[width=0.49\textwidth]{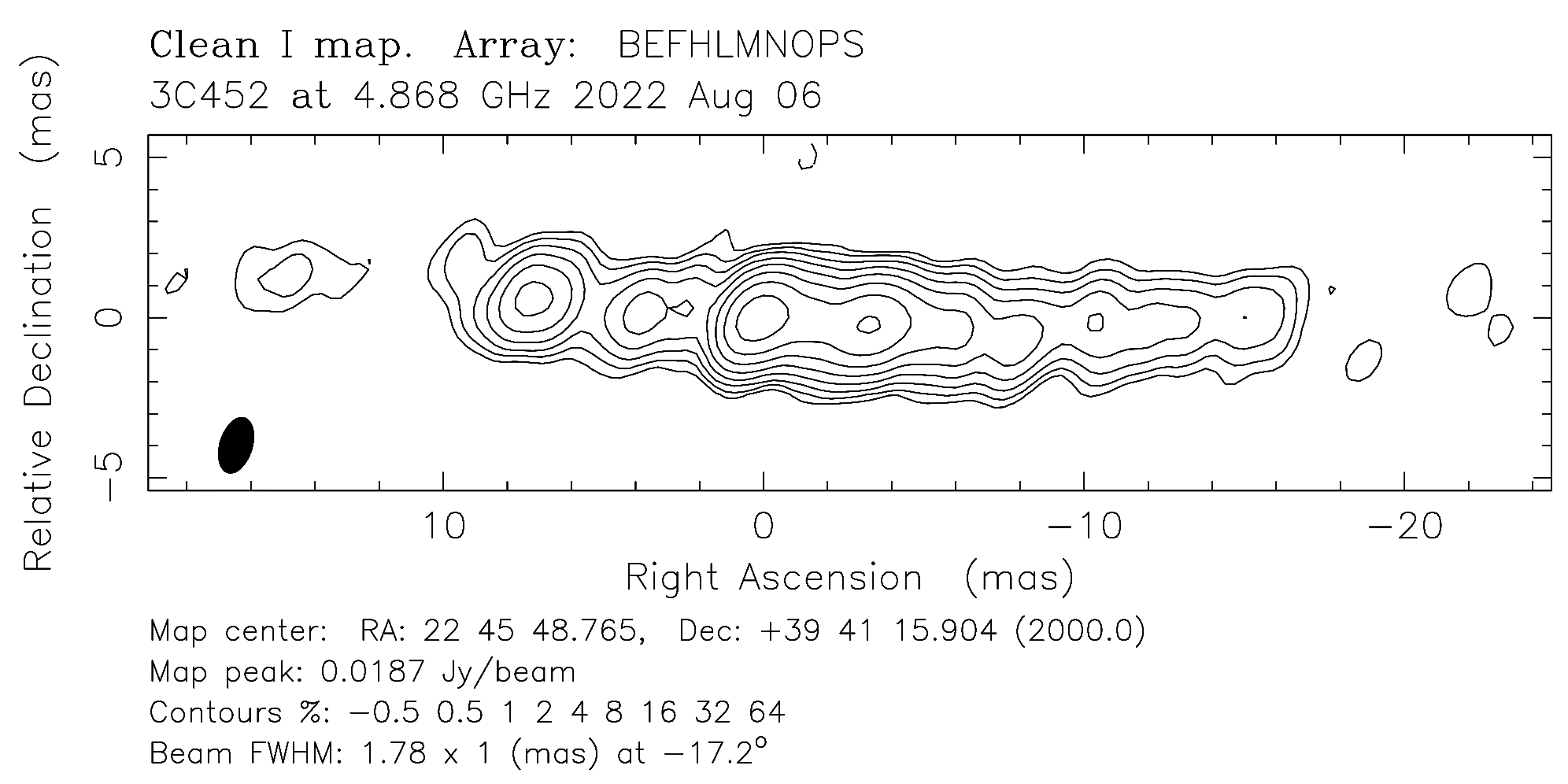}
         \includegraphics[width=0.49\textwidth]{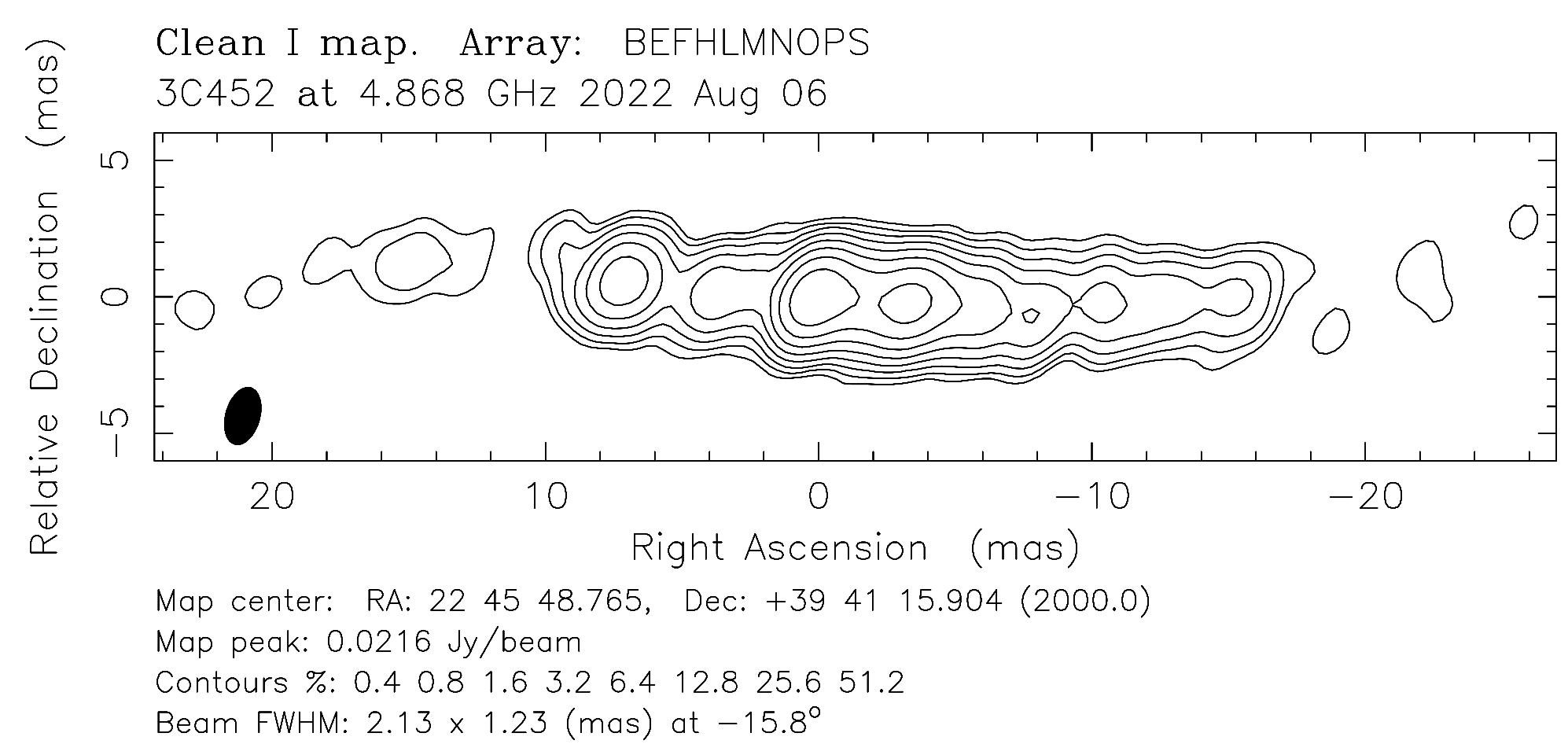}
        \includegraphics[width=0.49\textwidth]{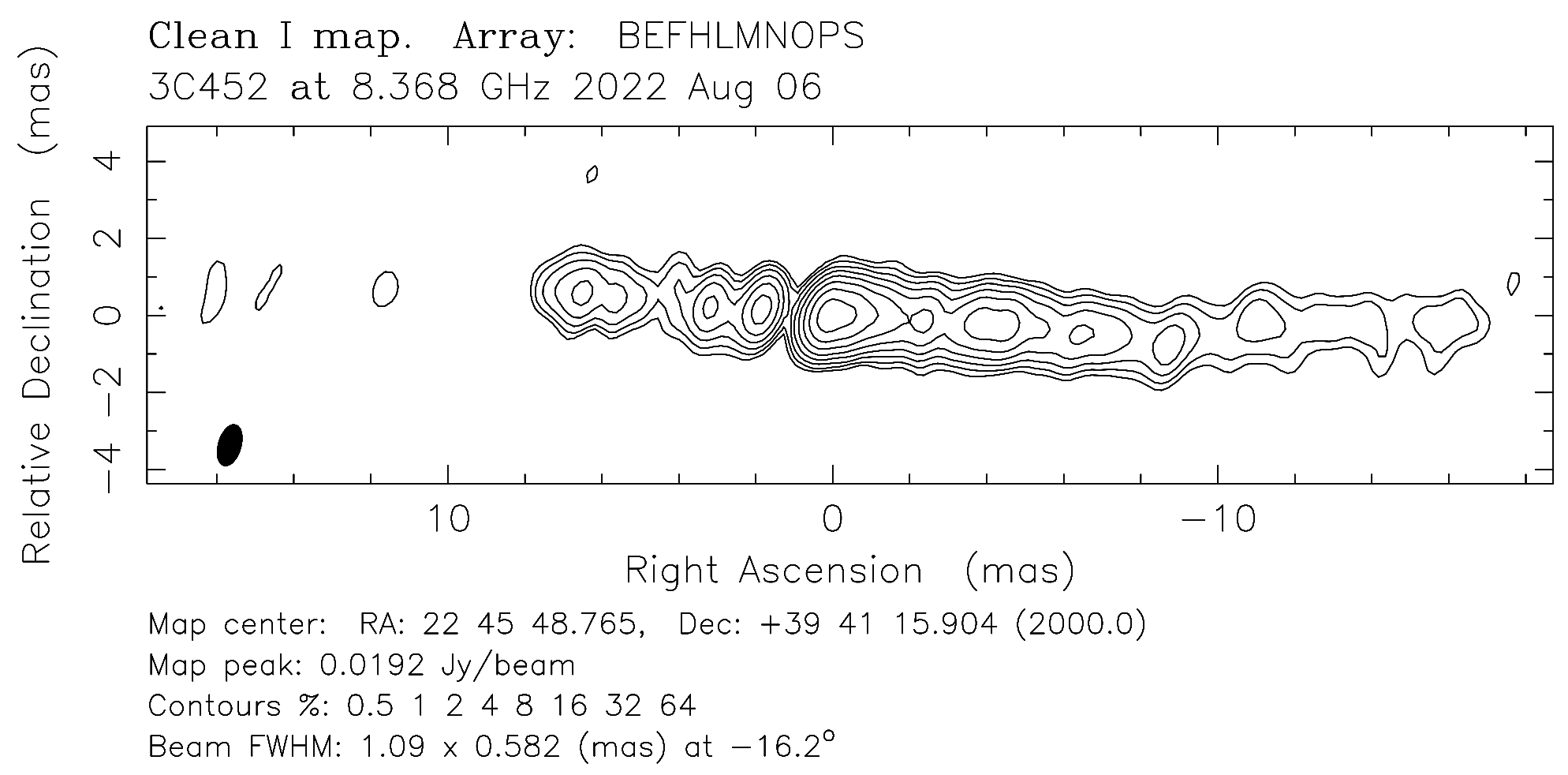}
        \includegraphics[width=0.49\textwidth]{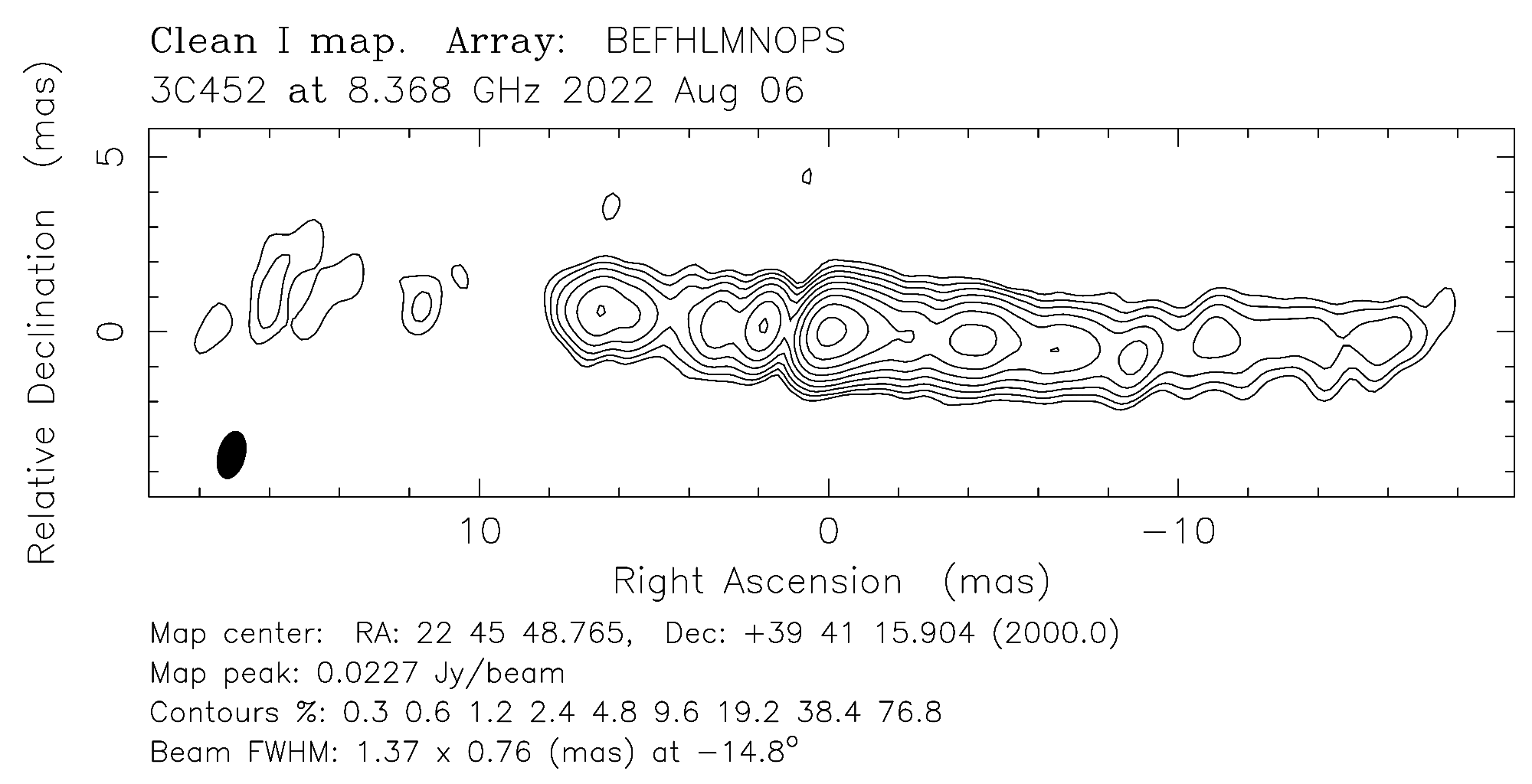}
        \includegraphics[width=0.49\textwidth]{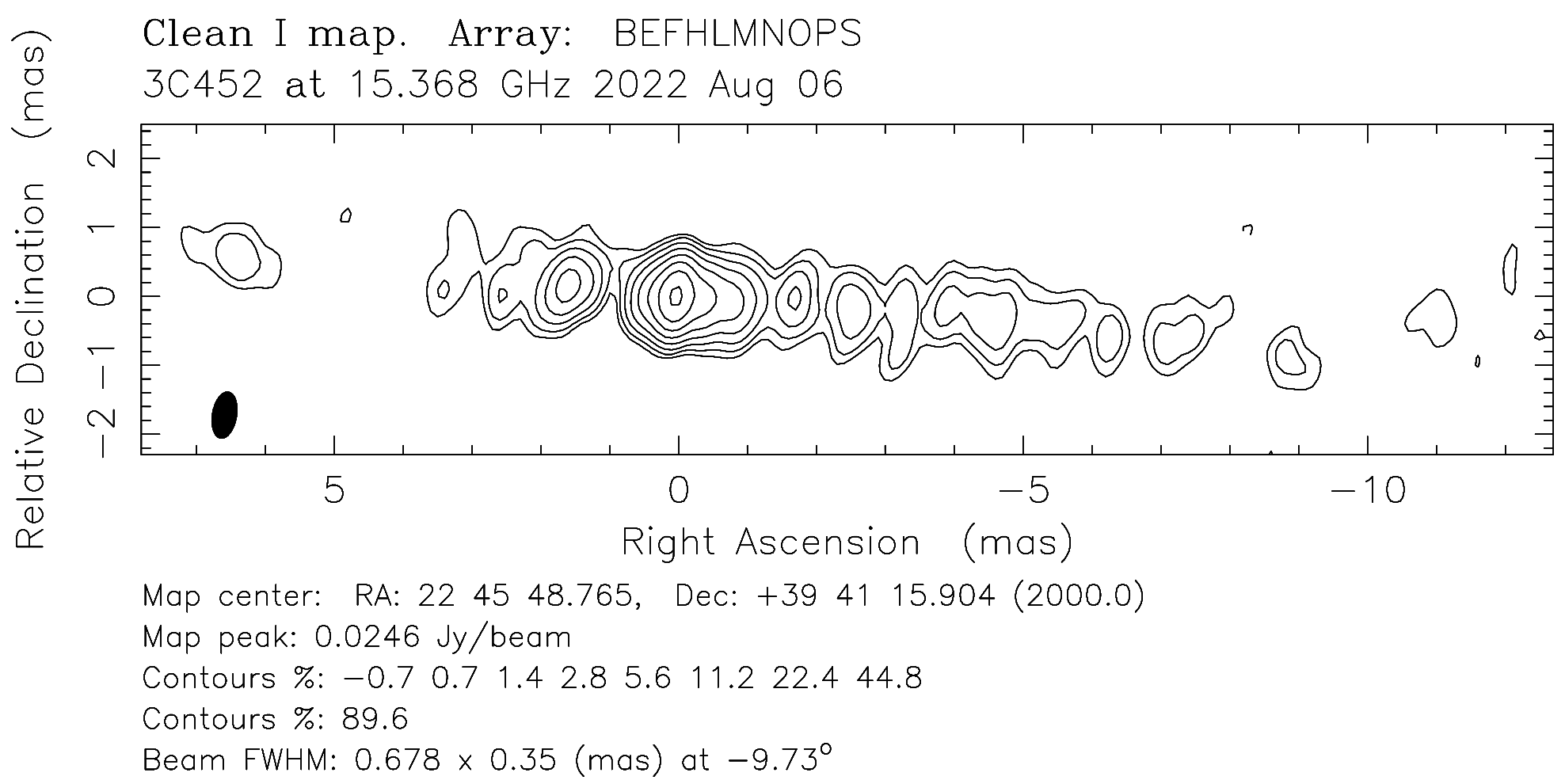}
        \includegraphics[width=0.49\textwidth]{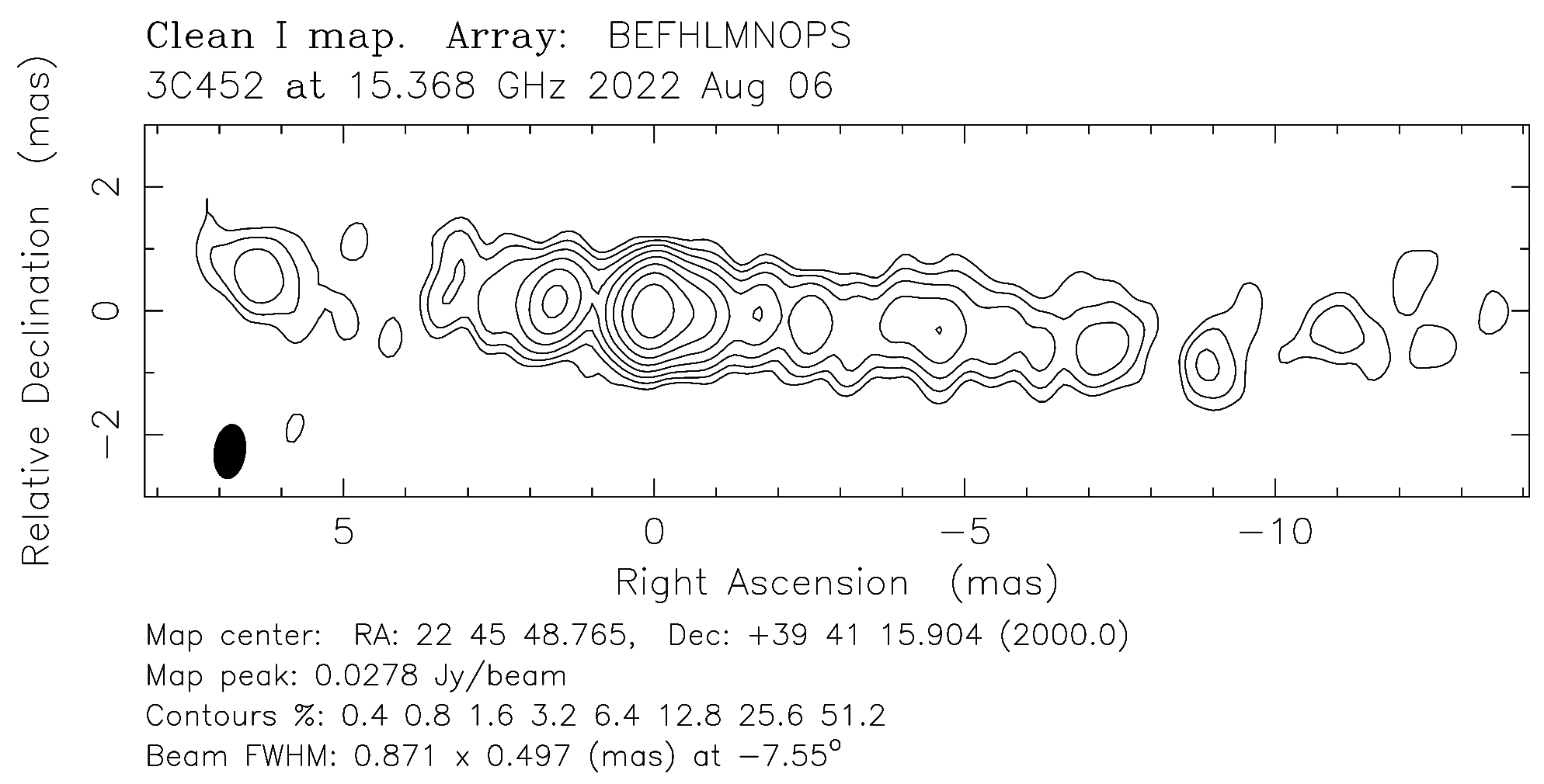}
        \includegraphics[width=0.49\textwidth]{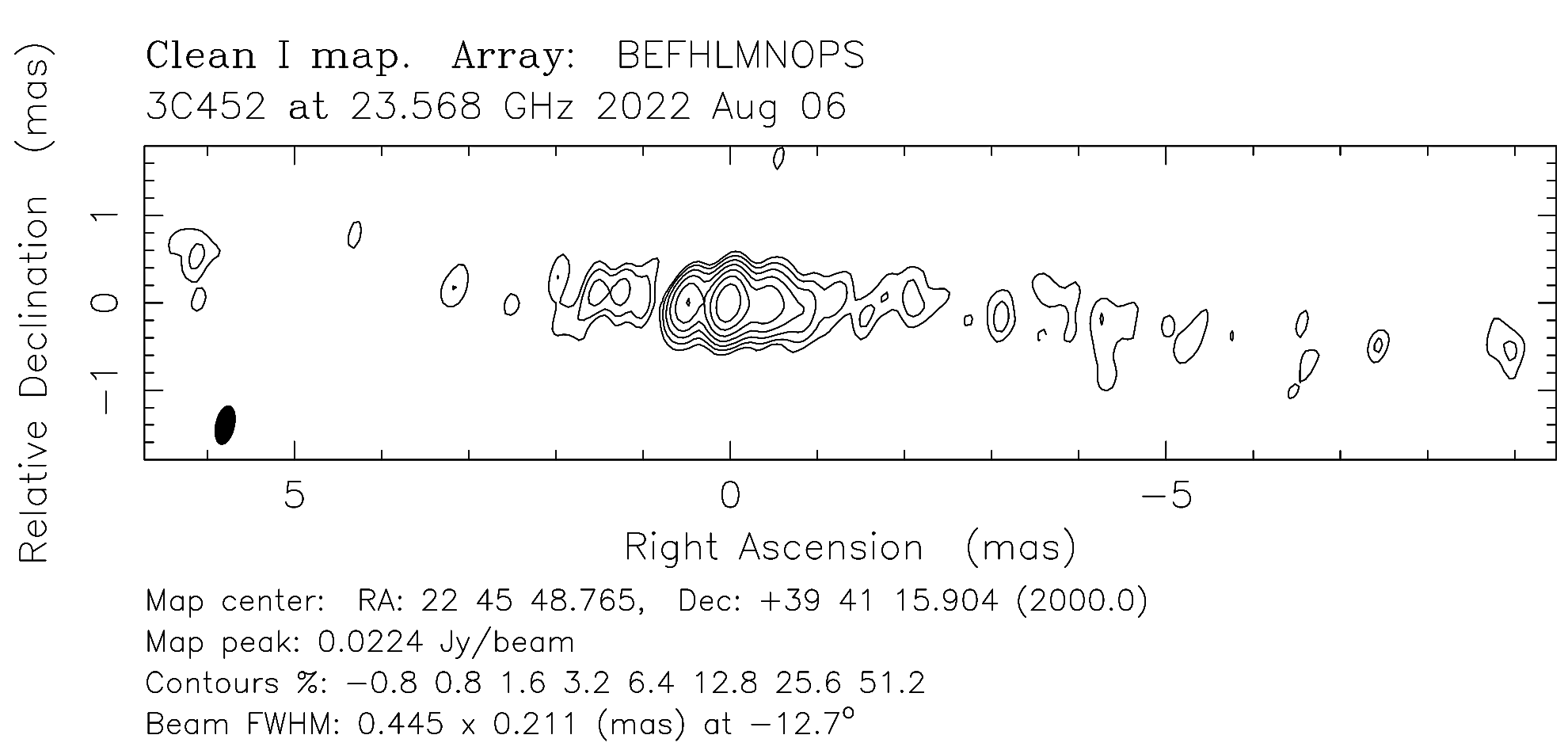}
        \includegraphics[width=0.49\textwidth]{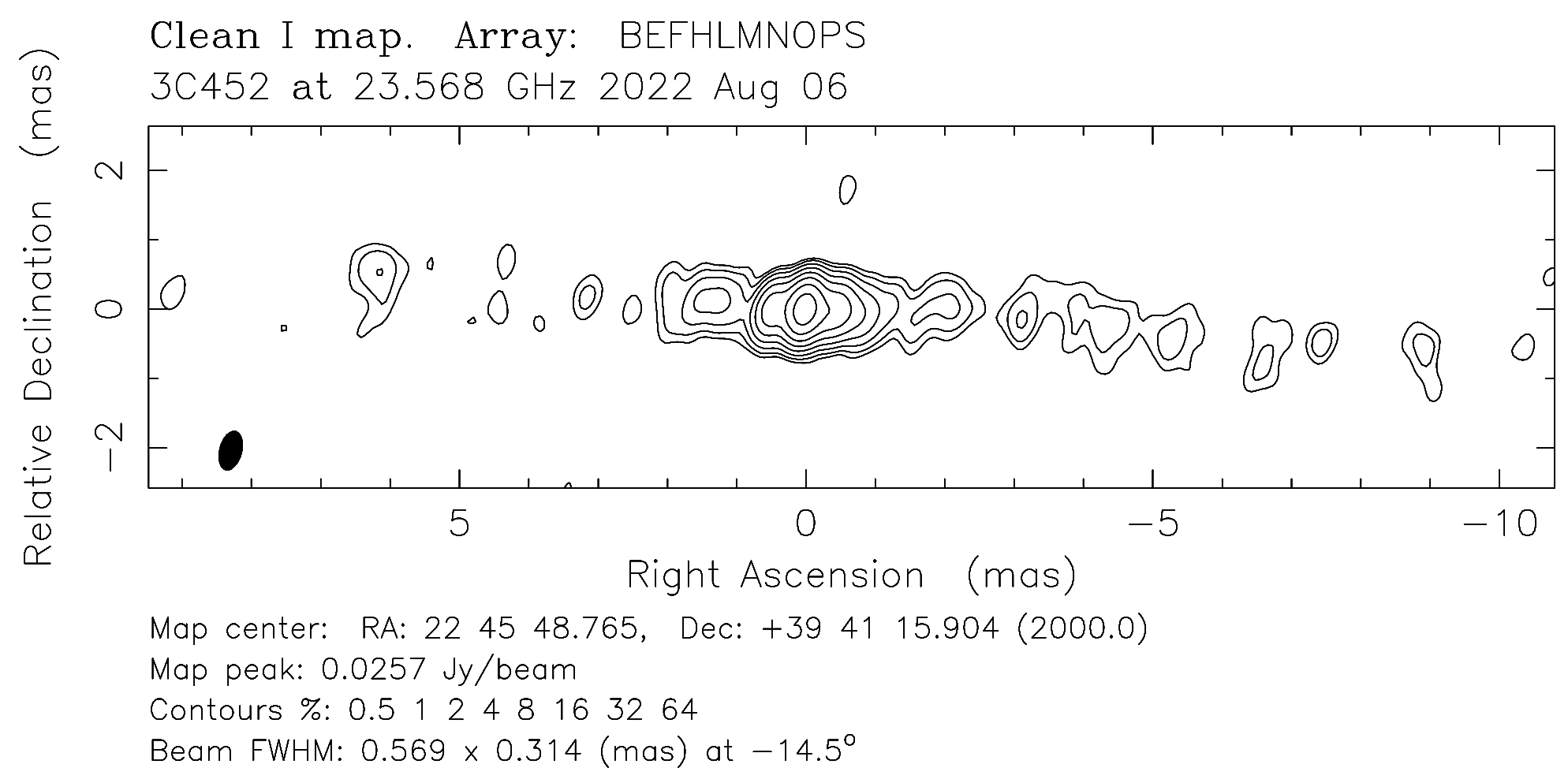}
        \includegraphics[width=0.43\textwidth]{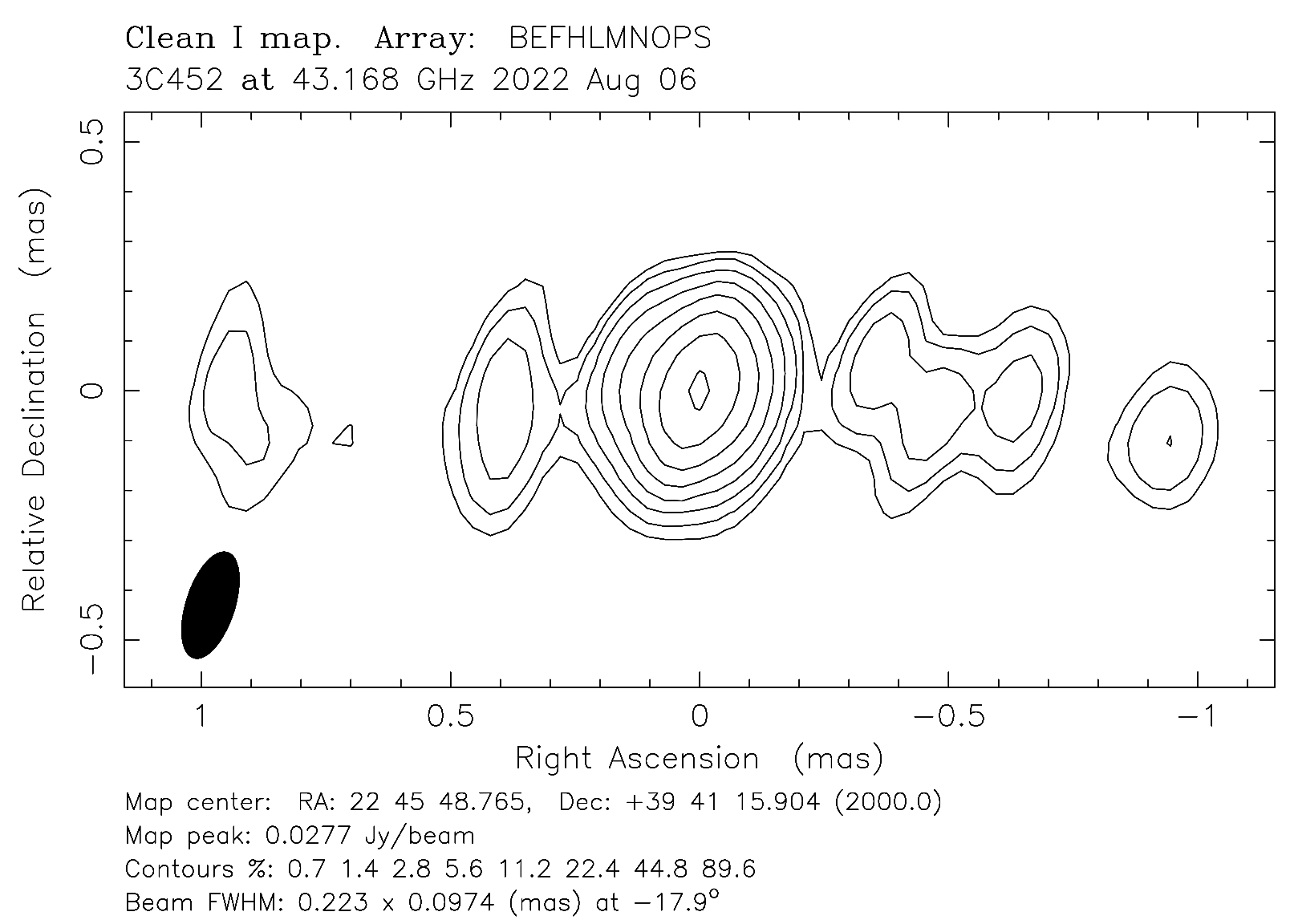}
            \includegraphics[width=0.43\textwidth]{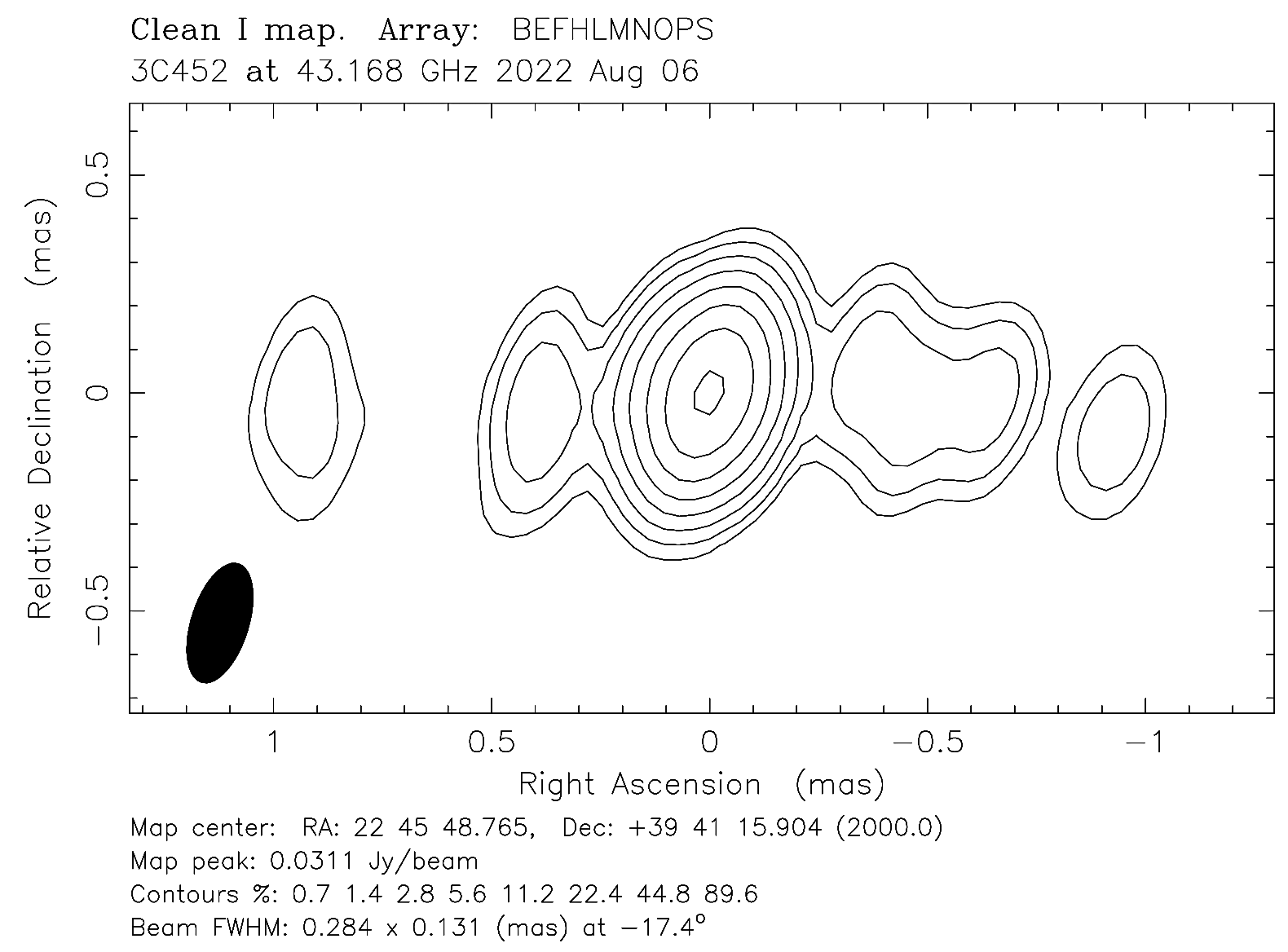}
       \caption{VLBI images of 3C 452 from project BM516B. From top to bottom: 4.9 GHz, 8.4 GHz, 15.4 GHz, 23.6 GHz, and 43.2 GHz. Left panels: images with uniform weighting. Right panels: images with natural weighting.}
    \label{fig:epoch-B-uniform}
\end{figure*}

\begin{table*}[!ht]
\centering
\footnotesize
\caption{Properties of the \texttt{MODELFIT} Gaussian components for the multi-frequency VLBI observations of 3C 452, from project code BM516A. }
\textbf{Project Code: BM516A}\\
    \label{tab:table A1}
\begin{tabular}{ccccccc}
\hline
Frequency      & $r$         & $S$          & $pa$   & FWHM               \\
 $\mathrm {[GHz]}$& $\mathrm {[mas]}$& $ \mathrm {[mJy]}$& $\rm{[deg]}$     & $\mathrm {[mas]}$  \\
\hline         

\hline

4.9 & 12.66 $\pm$ 0.72& 2.26 $\pm$ 0.79 & 85.6 $\pm$ 3.27& 4.42 $\pm$ 1.45\\ 
    & 6.97 $\pm$ 0.01& 7.57 $\pm$ 1.13 & 84.1 $\pm$ 0.11& 1.11 $\pm$ 0.03\\ 
    & 3.86 $\pm$ 0.03& 4.92 $\pm$ 0.99 & 86.7 $\pm$ 0.37& 2.05 $\pm$ 0.06\\ 
    & 0.00 $\pm$ 0.03& 26.53 $\pm$ 2.08 & -  & 0.97 $\pm$ 0.06\\ 
    & 1.67 $\pm$ 0.04& 16.17 $\pm$ 1.63 & -92.3 $\pm$ 1.34& 0.99 $\pm$ 0.08\\ 
    & 3.58 $\pm$ 0.04& 21.03 $\pm$ 1.87 & -93.3 $\pm$ 0.60& 1.08 $\pm$ 0.07\\ 
    & 5.63 $\pm$ 0.07& 9.16 $\pm$ 1.25 & -93.5 $\pm$ 0.67& 1.20 $\pm$ 0.13\\ 
    & 7.84 $\pm$ 0.07& 3.98 $\pm$ 0.82 & -95.2 $\pm$ 0.55& 0.94 $\pm$ 0.15\\ 
    & 10.55 $\pm$ 0.21& 4.61 $\pm$ 0.98 & -90.9 $\pm$ 1.14& 2.21 $\pm$ 0.42\\ 
    & 14.23 $\pm$ 0.31& 3.92 $\pm$ 0.97 & -90.2 $\pm$ 1.25& 2.73 $\pm$ 0.62\\ 
\hline
8.4 & 6.41 $\pm$ 0.08& 5.82 $\pm$ 1.14& 84.8 $\pm$ 0.69& 0.94 $\pm$ 0.15\\
    & 3.18 $\pm$ 0.02& 4.05 $\pm$ 0.92& 84.6 $\pm$ 0.31&  0.53 $\pm$ 0.03\\
    & 1.97 $\pm$ 0.03& 4.43 $\pm$ 0.96& 82.2 $\pm$ 0.89& 0.38 $\pm$ 0.06\\
    & 0.00 $\pm$ 0.02& 29.11 $\pm$ 2.45& - &  0.62 $\pm$ 0.04\\
    & 1.08 $\pm$ 0.05& 11.07 $\pm$ 1.55& -90.2 $\pm$ 2.79&  0.91 $\pm$ 0.10\\
    & 2.39 $\pm$ 0.01& 4.34 $\pm$ 0.94& -93.2 $\pm$ 0.26&  0.14 $\pm$ 0.02\\
    & 4.04 $\pm$ 0.06& 13.29 $\pm$ 1.75& -93.4 $\pm$ 0.89&  1.11 $\pm$ 0.12\\
    & 6.54 $\pm$ 0.01& 7.57 $\pm$ 1.36& -93.6 $\pm$ 0.13&  1.42 $\pm$ 0.03\\
    & 10.04 $\pm$ 0.02& 3.98 $\pm$ 0.99& -92.2 $\pm$ 0.12&  1.57 $\pm$ 0.04\\
    & 13.78 $\pm$ 0.03& 1.42 $\pm$ 0.56& -91.2 $\pm$ 0.12&  0.29 $\pm$ 0.06\\

\hline
15.4 & 6.29 $\pm$ 0.06& 2.11 $\pm$ 0.63& 83.1 $\pm$ 0.53  & 0.47 $\pm$ 0.12\\ 
     & 1.62 $\pm$ 0.02& 6.91 $\pm$ 1.08& 84.3 $\pm$ 0.68& 0.31 $\pm$ 0.04\\ 
     & 0.24 $\pm$ 0.01& 11.02 $\pm$ 1.34& 80.1 $\pm$ 2.39 & 0.22 $\pm$ 0.02\\ 
     & 0.00 $\pm$ 0.001& 19.75 $\pm$ 1.78& - & 0.15 $\pm$ 0.01\\ 
     & 0.37 $\pm$ 0.01& 10.46 $\pm$ 1.32& -94.3 $\pm$2.22 & 0.29 $\pm$ 0.03\\ 
     & 1.90 $\pm$ 0.08& 6.63 $\pm$ 1.22& -92.5 $\pm$ 2.31& 0.91 $\pm$ 0.15\\ 
     & 4.26 $\pm$ 0.06& 5.28 $\pm$ 1.05& -92.6 $\pm$ 0.88& 0.73 $\pm$ 0.13\\ 
     & 6.27 $\pm$ 0.14& 5.05 $\pm$ 1.13& -93.9 $\pm$ 1.24& 1.29 $\pm$ 0.27\\ 
\hline

23.6 & 5.80 $\pm$ 0.13& 0.95 $\pm$ 0.44& 84.0 $\pm$ 1.29& 0.66 $\pm$ 0.26\\ 
     & 3.19 $\pm$ 0.09& 1.96 $\pm$ 0.61& 86.3 $\pm$ 1.54& 0.65 $\pm$ 0.17 \\ 
     & 1.33 $\pm$ 0.03& 5.62 $\pm$ 0.98& 84.1 $\pm$ 1.20& 0.40 $\pm$ 0.06\\ 
     & 0.40 $\pm$ 0.01& 8.70 $\pm$ 1.19& 86.6 $\pm$ 1.22& 0.17 $\pm$ 0.02\\ 
     & 0.00 $\pm$ 0.001& 25.40 $\pm$ 2.02& - & 0.03 $\pm$ 0.002\\ 
     & 0.52 $\pm$ 0.01& 10.76 $\pm$ 1.33& -93.6 $\pm$ 1.46& 0.28 $\pm$ 0.03\\ 
     & 2.05 $\pm$ 0.05& 2.47 $\pm$ 0.66& -94.2 $\pm$ 1.22& 0.49 $\pm$ 0.11\\ 
     & 3.72 $\pm$ 0.09& 3.35 $\pm$ 0.83& -94.9 $\pm$ 1.33& 0.78 $\pm$ 0.17\\ 
     & 5.04 $\pm$ 0.02& 1.42 $\pm$ 0.49& -93.6 $\pm$ 0.20& 0.14 $\pm$ 0.03\\ 
     & 7.05 $\pm$ 0.04& 1.02 $\pm$ 0.45& -94.5 $\pm$ 0.30& 0.55 $\pm$ 0.07\\ 

\end{tabular}
\tablefoot{Col. 1: Frequency. Col. 2: Radial separation from the core. Col. 3: Integrated flux density. Col. 4: Position angle. Col. 5: Transverse size (FWHM).}
\end{table*}

\begin{table*}[!ht]
\centering
\footnotesize
\caption{Properties of the \texttt{MODELFIT} Gaussian components for the multi-frequency VLBI observations of 3C 452, from project code BM516B. }
\textbf{Project Code: BM516B}\\
 \label{tab:table A2}
\begin{tabular}{ccccccc}
\hline
Frequency      & $r$         & $S$          & $pa$   & FWHM               \\
$\mathrm {[GHz]}$& $\mathrm {[mas]}$& $\mathrm {[mJy]}$& $\rm{[deg]}$     & $\mathrm {[mas]}$  \\
\hline         

\hline
4.9 & 6.63 $\pm$ 0.08& 7.57 $\pm$ 1.11 & 85.4 $\pm$ 0.68& 1.25 $\pm$ 0.16\\
    & 3.15 $\pm$ 0.06& 4.49 $\pm$ 0.82& 85.4 $\pm$ 1.10& 0.82 $\pm$ 0.12\\
    & 0.00 $\pm$ 0.01& 17.10  $\pm$ 1.55& -  & 0.25 $\pm$ 0.02\\
    & 0.95 $\pm$ 0.02& 15.09 $\pm$ 1.46& -95.2 $\pm$ 0.96 & 0.44 $\pm$ 0.03\\
    & 2.35 $\pm$ 0.04& 11.78 $\pm$ 1.33& -91.7 $\pm$ 1.06 & 0.95 $\pm$ 0.09\\
    & 3.79 $\pm$ 0.03& 18.85 $\pm$ 1.67& -92.7 $\pm$ 0.52 & 0.96 $\pm$ 0.07\\
    & 5.74 $\pm$ 0.10& 12.49 $\pm$ 1.52& -92.7 $\pm$ 1.03 & 1.87 $\pm$ 0.21\\
    & 8.47 $\pm$ 0.05& 4.30 $\pm$ 0.79& -93.5 $\pm$ 0.33 & 0.68 $\pm$ 0.10\\
    & 11.17 $\pm$ 0.12& 3.50 $\pm$ 0.76& -91.5 $\pm$ 0.62 & 1.28 $\pm$ 0.24\\
    & 14.62 $\pm$ 0.08& 2.14 $\pm$ 0.57& -92.4 $\pm$ 0.30& 0.72 $\pm$ 0.15\\

\hline
        8.4  & 6.39 $\pm$ 0.07& 4.39 $\pm$ 0.84& 85.6 $\pm$ 0.62& 0.84 $\pm$ 0.14\\
             & 3.00 $\pm$ 0.07& 3.64 $\pm$ 0.77& 87.9 $\pm$ 1.30& 0.76 $\pm$ 0.14\\
             & 1.74 $\pm$ 0.01& 4.84 $\pm$ 0.83& 85.2 $\pm$ 0.24& 0.12 $\pm$ 0.01\\
             & 0.00 $\pm$ 0.01& 21.41 $\pm$ 1.74& - & 0.33 $\pm$ 0.02\\
             & 0.63 $\pm$ 0.01& 13.59 $\pm$ 1.39& -90.6 $\pm$ 1.04& 0.30 $\pm$ 0.02\\
             & 1.37 $\pm$ 0.03& 8.75 $\pm$ 1.14& -90.7 $\pm$ 1.34& 0.61 $\pm$ 0.06\\
             & 2.54 $\pm$ 0.04& 6.40 $\pm$ 0.99& -93.4 $\pm$ 0.98& 0.68 $\pm$ 0.09\\
             & 4.28 $\pm$ 0.05& 13.08 $\pm$ 1.51& -92.8 $\pm$ 0.74& 1.07 $\pm$ 0.11\\
             & 6.42 $\pm$ 0.10& 5.87 $\pm$ 1.04& -93.7 $\pm$ 0.87& 1.22 $\pm$ 0.19\\
             & 8.81 $\pm$ 0.12& 3.26 $\pm$ 0.78& -93.7 $\pm$ 0.77& 1.11 $\pm$ 0.24\\
             & 11.50 $\pm$ 0.07& 1.92 $\pm$ 0.55& -91.9 $\pm$ 0.34& 0.59 $\pm$ 0.14\\
             & 14.67 $\pm$ 0.29& 2.25 $\pm$ 0.71& -91.2 $\pm$ 1.14& 1.99 $\pm$ 0.58\\
        \hline
15.4    & 6.42 $\pm$ 0.03& 1.96 $\pm$ 0.52& 84.6 $\pm$ 0.30& 0.68 $\pm$ 0.07\\
        & 2.48 $\pm$ 0.12& 2.30 $\pm$ 0.62& 85.5 $\pm$ 2.87& 1.00 $\pm$ 0.25\\
        & 1.57 $\pm$ 0.02& 6.09 $\pm$ 0.84& 84.5 $\pm$ 0.66& 0.33 $\pm$ 0.04\\
        & 0.34 $\pm$ 0.01& 6.60 $\pm$ 0.86& 88.2 $\pm$ 2.19& 0.26 $\pm$ 0.03\\
        & 0.00 $\pm$ 0.003& 24.71 $\pm$ 1.65& - & 0.12 $\pm$ 0.01\\
        & 0.38 $\pm$ 0.01& 8.67 $\pm$ 0.99& -95.7 $\pm$ 1.71& 0.26 $\pm$ 0.02\\
        & 0.85 $\pm$ 0.02& 5.93 $\pm$ 0.83& -96.7 $\pm$ 1.53& 0.40 $\pm$ 0.04\\
        & 1.74 $\pm$ 0.01& 1.82 $\pm$ 0.46& -91.5 $\pm$ 0.49& 0.16 $\pm$ 0.03\\
        & 2.59 $\pm$ 0.06& 3.54 $\pm$ 0.70& -95.0 $\pm$ 1.37& 0.70 $\pm$ 0.12\\
        & 4.55 $\pm$ 0.07& 5.86 $\pm$ 0.97& -93.4 $\pm$ 0.90& 0.94 $\pm$ 0.14\\
        & 6.89 $\pm$ 0.14& 3.15 $\pm$ 0.78& -94.3 $\pm$ 1.20& 1.23 $\pm$ 0.29\\
        & 9.01 $\pm$ 0.04& 0.98 $\pm$ 0.34& -95.7 $\pm$ 0.29& 0.33 $\pm$ 0.09\\

\hline
23.6 & 5.08 $\pm$ 0.05& 3.82 $\pm$ 0.89& 85.0 $\pm$ 1.19& 2.30 $\pm$ 0.61\\ 
     & 1.34 $\pm$ 0.05& 3.65 $\pm$ 0.85& 86.1 $\pm$ 2.18& 0.47 $\pm$ 0.10\\ 
     & 0.47 $\pm$ 0.01& 7.34 $\pm$ 1.09& 89.2 $\pm$ 0.74& 0.11 $\pm$ 0.01\\ 
     & 0.00 $\pm$ 0.01& 29.60 $\pm$ 2.19& -  & 0.17 $\pm$ 0.01\\ 
     & 0.42 $\pm$ 0.01& 9.38 $\pm$ 1.25& -93.7 $\pm$ 1.77& 0.24 $\pm$ 0.03\\ 
     & 0.82 $\pm$ 0.03& 3.52 $\pm$ 0.83& -95.6 $\pm$ 2.36& 0.35 $\pm$ 0.07\\ 
     & 1.79 $\pm$ 0.08& 1.78 $\pm$ 0.61& -90.6 $\pm$ 2.54& 0.51 $\pm$ 0.15\\ 
     & 4.19 $\pm$ 0.16& 2.75 $\pm$ 0.96& -92.7 $\pm$ 2.17& 0.95 $\pm$ 0.32\\ 
     & 5.34 $\pm$ 0.05& 0.97 $\pm$ 0.43& -94.7 $\pm$ 0.57& 0.30 $\pm$ 0.11\\ 
     & 9.01 $\pm$ 0.04 & 0.86 $\pm$ 0.40& -94.4 $\pm$ 0.23& 0.20 $\pm$ 0.07\\ 
     
\hline
43.2  & 0.92 $\pm$ 0.01& 1.20 $\pm$ 0.52& 77.7 $\pm$ 0.43& 0.04 $\pm$ 0.01\\
        & 0.39 $\pm$  0.01& 3.20 $\pm$ 0.84& 88.9 $\pm$ 1.57& 0.10 $\pm$ 0.02\\
        & 0.00 $\pm$ 0.003& 42.25 $\pm$ 2.97& -  & 0.10 $\pm$ 0.01\\
        & 0.38 $\pm$ 0.01& 4.66 $\pm$ 1.06& -95.7 $\pm$ 2.23 & 0.15 $\pm$ 0.03\\
        & 0.90 $\pm$ 0.01& 2.13 $\pm$ 0.69& -94.9 $\pm$ 0.93& 0.11 $\pm$ 0.03\\
\end{tabular}
\tablefoot{Col. 1: Frequency. Col. 2: Radial separation from the core. Col. 3: Integrated flux density. Col. 4: Position angle. Col. 5: Transverse size (FWHM).}
\end{table*}

\begin{figure*}[!h]   \includegraphics[width=1\textwidth] 
{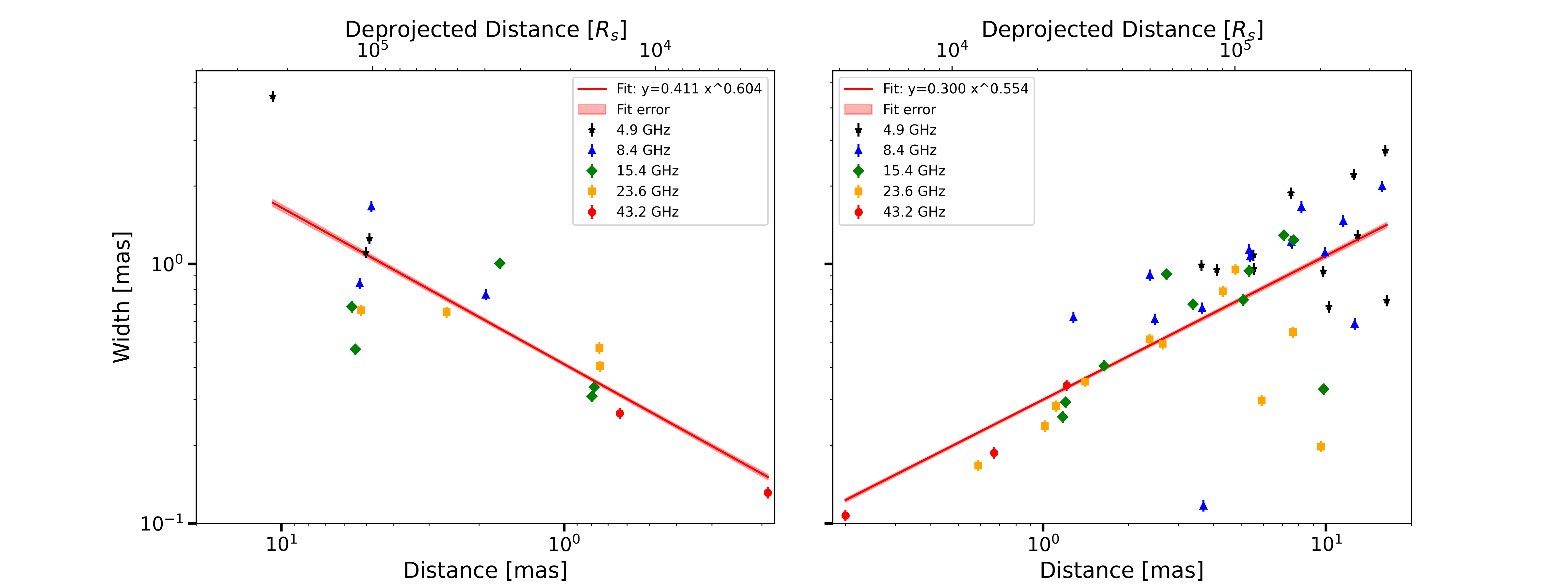}
       \caption{Collimation profile of the receding counter-jet (left) and approaching jet (right) derived from \texttt{MODELFIT} components at all frequencies and epochs.
        Only fully resolved components located at distances larger than one beam from the core and with FWHM larger than half the beam minor axis are included. 
        The fitted power-law indices are consistent with those obtained from the pixel-based analysis (see Fig.~\ref{fig:stacked images-collimation profiles}), confirming the parabolic expansion of both jets. The shaded Fit Error band is included in the legend but is very small compared to the data scatter and thus not easily visible in the plot.
    }
        \label{fig:collimation-modelfit}
\end{figure*}

\begin{table*}[!ht]
    \centering
    \caption{\texttt{MODELFIT} core component and map parameters used to derive the core resolution limit and brightness temperature at all frequencies for BM516A and BM516B.}
    \small
     \label{tab:table A3}
    \begin{tabular}{ccccccccc}
        \hline
        P.C. & Frequency & $S^{\mathrm{fit}}_{\mathrm{p}}$  & $\sigma_{\mathrm{fit}}$ & $b_{\mathrm{maj}}$ & $b_{\mathrm{min}}$ & $d_{\mathrm{lim}}$ &  $d/d_{\mathrm{lim}}$ & $ T_{\mathrm{B}}^{\mathrm{c}}$ \\
        & [GHz] & {[mJy/beam]} & {[mJy/beam]} & {[mas]} & {[mas]}  & {[mas]} & & [10$^{10}$ K] \\
        \hline
        BM516A & 4.9 & 21.75 $\pm$ 1.32 & 0.08 & 2.57 & 1.77 & 0.122 & 7.97 & 0.16 $\pm$ 0.02 \\
               & 8.4 & 23.24 $\pm$ 1.53 & 0.10 & 1.58 & 0.88 & 0.073& 8.51 & 0.14 $\pm$ 0.02 \\
               & 15.4 & 18.46 $\pm$ 1.22 & 0.08 & 0.71 & 0.47 & 0.036& 4.20 & 0.49 $\pm$ 0.08 \\
               & 23.6 & 25.48 $\pm$ 1.43 & 0.08 & 0.66 & 0.56 & 0.038& 0.79 & $>4.17$ \\
        \hline
        BM516B & 4.9 &  16.02 $\pm$ 1.06 & 0.07 & 2.13 & 1.23 & 0.101& 2.48 & 1.50 $\pm$ 0.25 \\
               & 8.4 & 18.85 $\pm$ 1.15 & 0.07 & 1.37 & 0.76 & 0.058& 5.64 & 0.37 $\pm$ 0.05 \\
               & 15.4 & 23.48 $\pm$ 1.14 & 0.05 & 0.87 & 0.50 & 0.030& 4.01 & 0.95 $\pm$ 0.11 \\
               & 23.6 & 25.20 $\pm$ 1.42 & 0.08 & 0.56 & 0.31 & 0.022& 7.59 & 0.24 $\pm$ 0.03 \\
               & 43.2 & 31.66 $\pm$ 1.78 & 0.10 & 0.28 & 0.13 & 0.010& 9.81 & 0.30 $\pm$ 0.04 \\
        \hline
    \end{tabular}
    \tablefoot{Col. 1: Project code. Col. 2: Observing frequency. Col. 3: Peak intensity of the \texttt{MODELFIT} core component and associated uncertainty. Col. 4:  Post-fit rms of the \texttt{MODELFIT} residual map. Col. 5: Beam major axis. Col. 6: Beam minor axis. Col. 7: Resolution limit. Col. 8: Ratio between the core size and the resolution limit. Col. 9: Core brightness temperature and associated uncertainty. A lower limit is given when the core size is smaller than the resolution limit.}
\end{table*}

\end{appendix}

\end{document}